\documentclass[draft]{agujournal2018}
\usepackage{xcolor}

\usepackage{apacite}
\usepackage{url} 
\justifying
\usepackage{amsmath}

\draftfalse

\journalname{Journal of Advances in Modeling Earth Systems (JAMES)}

\begin{document}

%
%


\title{A new hybrid mass-flux/ high-order turbulence closure for ocean vertical mixing}

%
%




\authors{Amrapalli Garanaik\affil{1}, Filipe Pereira\affil{2}, Katherine Smith\affil{2}, Rachel Robey\affil{3}, Qing Li\affil{4},  Brodie Pearson\affil{1}, and Luke Van Roekel\affil{2}}


\affiliation{1}{College of Earth, Ocean, and Atmospheric Sciences, Oregon State University, Corvallis, OR, USA}
\affiliation{2}{Fluid Dynamics and Solid Mechanics, Los Alamos National Laboratory, Los Alamos, NM, USA}
\affiliation{3}{Department of Applied Mathematics,  University of Colorado Boulder, Boulder, CO, USA}
\affiliation{4}{Earth, Ocean and Atmospheric Sciences Thrust, The Hong Kong University of Science and Technology (Guangzhou), Guangzhou, Guangdong, China}




\correspondingauthor{Amrapalli Garanaik}{amrapalli.garanaik@oregonstate.edu}




\begin{keypoints}
\item A new physically-motivated, PDF-based parameterization of ocean surface boundary layer turbulence is  presented.
\item The non-local forcing is included naturally and the scheme provides a closed set of equations with realizable closure assumptions.
\item The mixing scheme performs well for convective turbulence across different vertical resolutions.
\end{keypoints}

%
%


\begin{abstract}

While various parameterizations of vertical turbulent fluxes at different levels of complexity have been proposed, each has its own limitations. For example, simple first-order closure schemes such as the K-Profile Parameterization (KPP) lack energetic constraints; two-equation models like $k-\varepsilon$ directly solve an equation for the turbulent kinetic energy but do not account for non-local fluxes, and high-order closures that include the non-local transport terms are computationally expensive. To address these, here we extend the Assumed-Distribution Higher-Order Closure (ADC) framework originally proposed for the atmospheric boundary layer and apply it to the ocean surface boundary layer (OSBL). By assuming a probability distribution function relationship between the vertical velocity and tracers, all second-order and higher-order moments are exactly constructed and turbulence closure is achieved in the ADC scheme. In addition, the ADC parameterization scheme has full energetic constraints. We have tested the ADC scheme against a combination of large eddy simulation (LES), KPP, and $k-\varepsilon$ for surface buoyancy-driven convective mixing and found that the ADC scheme is robust with different vertical resolutions and compares well to the LES results.
\end{abstract}

\section*{Plain Language Summary}

The upper ocean (order of few tens of meters depth from the surface) has a substantial influence on our climate and weather system. Specifically, upper ocean mixing processes play a key role in modulating global heat budget in the ocean and atmosphere by mixing heat deeper into the ocean or warming the atmosphere above. Accurate representation of the effects of these mixing processes in the global climate and ocean models is crucial for understanding our current and changing climate. However, current mixing schemes used in these models have shown significant biases. Here, we develop a new physically-motivated mixing scheme for the upper ocean inspired from atmospheric mixing schemes. Results show that the proposed mixing scheme can simulate upper ocean mixing efficiently, suggesting its potential use in climate and ocean models to help reduce model biases. 

\section{Introduction}
The turbulent ocean surface boundary layer (OSBL) plays a key role in the evolution of the earth system and global ocean heat budget by communicating heat, mass, and momentum between the atmosphere and ocean interior \cite{Brainerd1995, Kantha2000}. The dynamics of the OSBL, which lead to temporal and spatial variability of quantities such as the OSBL depth and sea surface temperature (SST), are generally modulated by small-scale vertical turbulent fluxes. These small-scale turbulent motions are not resolved in general circulation models (GCMs) that are used to study the climate. Instead, the effects of vertical turbulent fluxes of heat, mass, and momentum are parameterized in GCMs through a vertical mixing scheme, e.g., bulk model \cite{Niiler1977}, $k-\varepsilon$ \cite{Rodi1987}, local K-theory \cite{Stull1988}, and K-Profile Parameterization \cite<KPP,>[]{Large1994}. The fidelity of these schemes is paramount to an accurate ocean simulation \cite{Eric2014}. However, existing parameterization schemes within GCMs have shown significant biases in representing both OSBL depth and SST \cite{Burchard2001, Kemper2011, Belcher2012,Li2019,Damerell2020}. These biases suggest that the parameterization of small-scale mixing in the OSBL is still an outstanding problem in modeling large-scale ocean dynamics \cite{Baylor2019}. 

\begin{figure}[h!]
\centering
\includegraphics[width=0.6\textwidth]{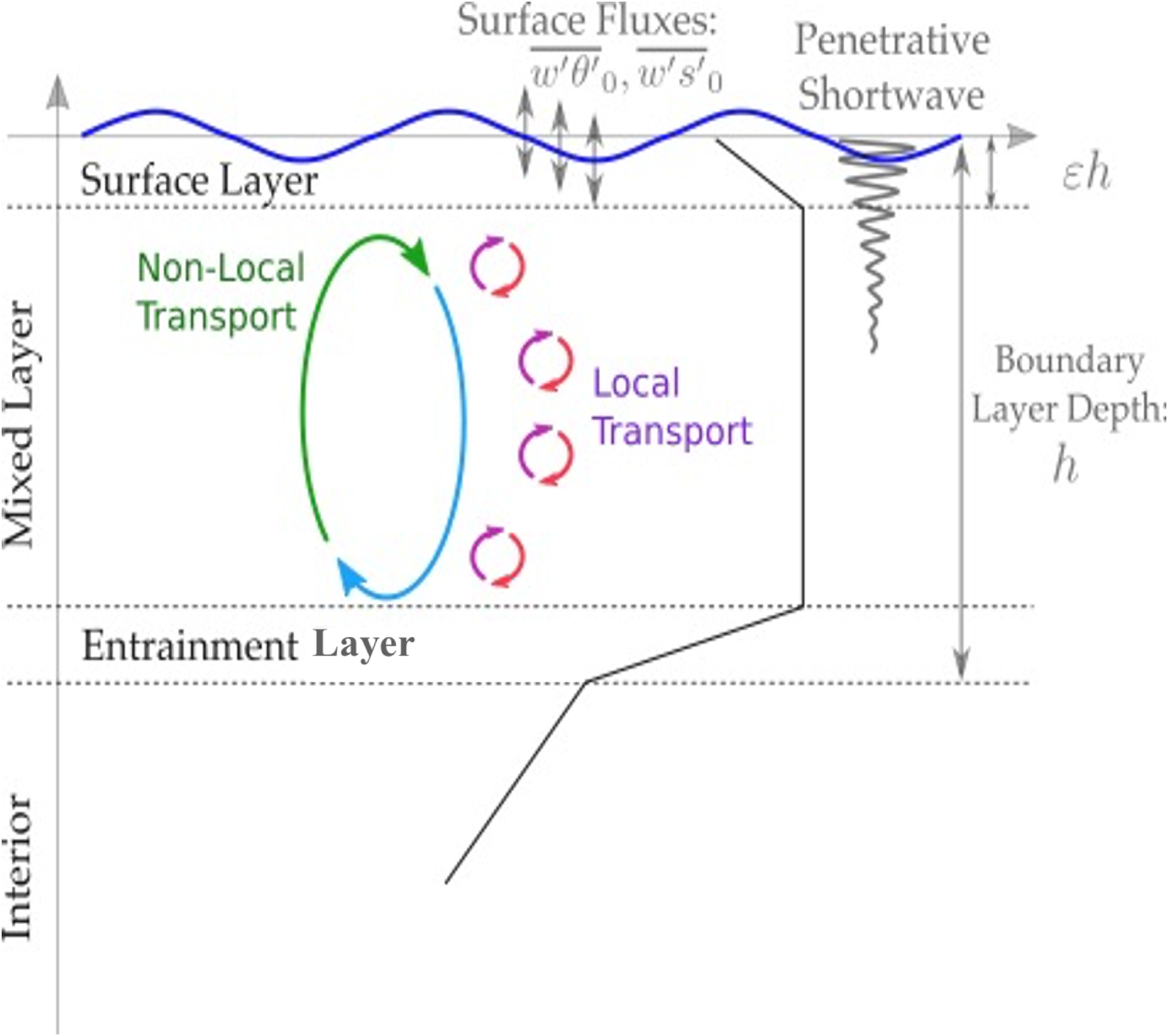}
\caption{Diagram of the structure of the OSBL and relevant forcing. The black profile shows an idealized potential temperature profile.}
\label{fig:osbl}
\end{figure}

Studies of the OSBL often invoke a boundary layer assumption under which the problem is reduced to a one-dimensional vertical column where vertical fluxes must be diagnosed and horizontal advection is ignored. Figure \ref{fig:osbl} illustrates the typical vertical OSBL structure. The OSBL has a depth $h$ and can typically be separated into three layers. First, the mixed layer is a region of strong turbulent mixing with vertically homogeneous properties that fills the bulk of the OSBL. Second, the surface layer, that lies between the mixed layer and the ocean surface, and is typically shallow with depth $\varepsilon h$ (where $\varepsilon\approx 0.1$). Finally, the entrainment layer lies between the mixed layer and the ocean interior or thermocline. As denoted in Figure \ref{fig:osbl}, the sources of mixing differ between layers and can generally be separated into `local' mixing driven by small-scale turbulence, and `non-local' mixing driven by large-scale turbulent structures, which transport properties long distances producing fluxes that are independent of local gradients. In the surface and entrainment layers, buoyancy gradients are large and most mixing is local, because turbulent motions are constrained to be small by the proximity to the surface and strong stratification respectively. Non-local mixing is also observed in the entrainment layer due to penetrative convective cells overshooting the thermocline. In the mixed layer, large eddies (e.g. convective plumes) can create significant non-local mixing in addition to local mixing resulting from smaller-scale turbulence. Any vertical mixing scheme must therefore strive to accurately parameterize both local and non-local contributions to the turbulent fluxes.

Traditional OSBL parameterization schemes do not typically account for both local and non-local mixing. For example, in the local K-theory \cite{Stull1988}, 
only prognostic equations for the mean ($\overline{\psi}$) are solved and turbulent fluxes ($
F_\psi = \ \overline{w'\psi'}$) are parameterized with an eddy diffusivity, $\kappa$. Neglecting all higher-order moment terms (that account for non-local turbulent fluxes), turbulent fluxes are defined to be proportional to the local gradient of the mean ($F_\psi=-\kappa \partial \overline{\psi}/\partial z$). Conversely, the plume type mass-flux closures \cite<MFC,>[]{Turner1986, Arakawa1969} developed for deep-convection parameterizations, parameterize the vertical turbulent fluxes in terms of a convective mass flux only and do not account for small-scale diffusive mixing \cite{Randall1992, Canuto2007}. The simple energetics-based bulk mixed layer models \cite{Kraus1967, Chen1994} assume a homogeneous boundary layer with infinite diffusivity that does not account for any internal structures of the OSBL. 
Because both local diffusion and non-local convection are important mixing mechanisms, attempts have been made over recent decades to develop unified parameterization schemes for both the atmosphere and ocean based on these conventional models \cite{Deardorff1972, Troen1986, Holtslag1993, Large1994, Siebesma1995, Lappen2001a, Canuto2007, Reichl2018}. In particular, \citet{Large1994} developed the KPP scheme for the OSBL by adding the effects of non-local convection to the local K-theory through a non-local term $\gamma$ such that turbulent fluxes of scalars become $F_\psi=-\kappa (\partial \overline{\psi}/\partial z + \gamma)$. KPP is the most widely used first-order vertical mixing scheme in GCMs that includes non-local mixing. However, many studies have discussed biases in KPP due to lack of energetics and its sensitivity to vertical resolutions and numerical implementations \cite{Burchard2001, Canuto2007, vanRoekel2018,Li2019, Zhu2019, Souza2020, Chor2021}.

In order to include the non-local transport terms, the time evolution equations of turbulent fluxes ($\overline{w’\psi’}$; a second-order moment) need to be solved without first-order approximation. However, this increases the number of prognostic equations making it difficult to solve in a GCM \cite{Burchard2001,MY1982, Cheng2006}. Therefore, various simplified second-order turbulent closure models have been proposed and implemented in GCMs. The most commonly used turbulent closure schemes of this type are either one-equation models, such as $1.5k$ \cite{Gaspar1990} and $2.5k$ \cite{MY1982}, or two-equation models, such as $k-\varepsilon$ \cite{Rodi1987}, $k-\omega$ \cite{Umlauf2003}, and generic length scale models \cite{UmlaufB2003}, where $k$ is the turbulent kinetic energy, $\varepsilon$ is the dissipation rate of $k$, and $\omega\propto\varepsilon/k$ is the specific dissipation rate. These vertical mixing schemes parameterize the eddy diffusivity by solving the prognostic equations for either $k$ or for both $k$ and $\varepsilon$. However, these closures still do not capture the non-local flux by convective turbulence, motivating the need for  higher-order closures \cite<HOC,>[]{Kantha2000, Canuto2007} which solve prognostic equations for at least the second-order turbulent moments, with some HOC schemes also solving for third- or higher-order moments \cite{Andre1978}. The latter HOC schemes explicitly evolve the higher-order transport terms which directly contribute to the non-local fluxes. The difficulty of moving to a HOC lies in the fact that prognostic equations for $n$-order moments always contain $(n+1)$-order moments. So while using a suitable HOC is more accurate than a lower-order closure, HOCs become computationally intractable in GCMs due to the increasing number of prognostic and diagnostic equations and the accompanying necessarily short integration time scales. HOC methods have indeed been developed and implemented with success \cite{Andre1978,Moeng1989,Cheng2005, Canuto2007}, however, they typically see less usage for reasons of complexity and added computational burden from the cascading size of the equation set. Another pitfall of HOC is that there is no guarantee that all the higher-order moments will be consistent with each other \cite{Andre1978}. To remedy these issues, assumed probability density functions (PDFs) are suggested for turbulent flows \cite{O'Brien1980}, such that all higher-order moments in HOC equations can be diagnosed from an assumed PDF of turbulent fluctuations without any realizability issues and with a reduced number of prognostic equations.

Based on assumed PDFs, \citet{Lappen2001a,Lappen2001c,Lappen2001b} developed a new mixing scheme by unifying HOC and MFC for the atmospheric boundary layer. This new assumed distribution closure (ADC) retains the HOC-physics and captures both local and non-local mixing in an energetically consistent manner. The PDF/mass-flux insight allows the diagnosis of high-order moments resulting from turbulent advection of lower-order moments, eliminating realizability issues and closing the second-order moment budgets with fewer prognostic equations than that required for HOC \cite{Lappen2001a, Golaz2002}. This ADC-unified HOC and MFC has been tested in the atmospheric boundary layer turbulence. However, the utility of ADC scheme has not been explored for OSBL turbulence. We note that some recent studies have used the MFC concept (non-local mixing) to study oceanic deep convection by modifying a plume model \cite{Canuto2007} and implementing the eddy-diffusivity-mass-flux (EDMF) closure \cite{Giordani2020}, however, neither of theses OSBL models have been tested under general oceanographic conditions other than sporadic deep-convection.

There is a clear need for a more reliable and plausible vertical mixing scheme for the OSBL, that includes the physics of local and non-local mixing, and is energetically consistent. In this paper, we present a new ADC mixing scheme for the OSBL, modified from the atmospheric ADC \citep{Lappen2001a,Lappen2001c, Lappen2001b}. To the authors' knowledge, this is the first time an assumed distribution higher-order closure scheme has been implemented in the OSBL. We've implemented the ADC model within the Model for Prediction Across Scales-Ocean  \cite<MPAS-Ocean,>[]{Ringler2013,Petersen2018}, the ocean component of the U.S. Department of Energy's Energy Exascale Earth System Model \cite<E3SM,>[]{Golaz2019} and evaluated this scheme against large eddy simulations (LES) of convective turbulence. The scheme will be evaluated under more general oceanographic conditions including wind- and wave-driven ocean mixing in companion papers. In what follows, we describe the new vertical mixing scheme, including the physical justification that underlies mass-flux models in section \ref{sec:methods}. In section \ref{sec:testcases}, we describe the LES model and single column test cases used to validate the new scheme. Results showing the evolution of the mean and higher-order moments are presented in section \ref{sec:results}. Finally, the discussion and concluding remarks are presented in section \ref{sec:diss_n_concl} and section \ref{sec:concl}, respectively.

\section{The Assumed Distribution Closure (ADC) Scheme}
\label{sec:methods}
A primary goal of OSBL parameterizations in GCMs is to estimate the vertical fluxes of heat and salt ($F_{\psi}$ where $\psi$ represents $\theta$ or $s$ respectively) because these tracer fluxes affect the evolution of the mean state in the following way:
\begin{linenomath*}
\begin{equation}
\frac{\partial \overline{\theta}}{\partial t}\approx-\frac{\partial F_{\theta}}{\partial z}+\overline{S_{\theta}} \qquad \textrm{and} \qquad \frac{\partial \overline{s}}{\partial t}\approx-\frac{\partial F_{s}}{\partial z}+\overline{S_{s}}.
\label{eq:GCM_equation}
\end{equation}
\end{linenomath*}
In this equation, overbars represent a spatio-temporal average over the grid-cell volume and time step of the GCM, $F_{\psi}$ represents the vertical flux of $\psi$ due to turbulent motions within the grid cell, and $S_{\psi}$ are sources and sinks of $\overline{\psi}$.

The new ocean ADC scheme emulates the vertical fluxes of heat and salinity in the upper ocean by evolving prognostic equations for several turbulence statistics, and combining these statistics to form a closed set of equations. For the present convective mixing comparisons, the key prognostic equations of the ADC scheme are those of four plume-scale properties: the tracer fluxes ($\overline{w'\theta'}$ and $\overline{w's'}$), the vertical turbulent kinetic energy ($\overline{w'^2}$), and the third-order moment of the vertical velocity ($\overline{w'^3}$). The equations for each of these are
\begin{linenomath*}
\begin{eqnarray}
   \label{eq_wtheta_hoc}
  \frac{\partial\overline{w'\theta'}}{\partial t}&=&
 -\underbrace{\frac{\partial\overline{w'^2\theta'}}{\partial z}}_\mathrm{Transport}
 -\underbrace{\overline{w'^2}\frac{\partial\overline{\theta}}{\partial z}
 + g(\alpha_T\overline{\theta'^2}-\beta_s\overline{\theta's'})}_\mathrm{Production}
  -\underbrace{\frac{1}{\rho_0}\overline{\theta'\frac{\partial p'}{\partial z}}}_\mathrm{Pressure}
  -\underbrace{\varepsilon_{w\theta}}_\mathrm{Dissipation} + \underbrace{SPS^{w\theta}}_\mathrm{Sub-plume},\\
  \nonumber \\ 
  \frac{\partial\overline{w's'}}{\partial t}&=&
 -\frac{\partial\overline{w'^2s'}}{\partial z}
 -\overline{w'^2}\frac{\partial\overline{s}}{\partial z}
 + g(\alpha_T\overline{s'\theta'}-\beta_s\overline{s'^2})
  -\frac{1}{\rho_0}\overline{s'\frac{\partial p'}{\partial z}}
  -\varepsilon_{ws}
  +SPS^{ws},\\
  \nonumber \\ 
  \frac{\partial\overline{w'^2}}{\partial t}&=&
  \frac{\partial\overline{w'^3}}{\partial z}
  +2\overline{w'b'}
  -2\overline{w'\frac{\partial p'}{\partial z}}
  -\varepsilon_{ww}
  +SPS^{ww},\\ 
  \nonumber \\ 
  \frac{\partial\overline{w'^3}}{\partial t}&=&
  -\frac{\partial\overline{w'^4}}{\partial z}
  +3\overline{w'^2}\frac{\partial\overline{w'^2}}{\partial z}
  +3\overline{w'^2b'}
  -3\overline{\bigg(w'^2\frac{\partial p'}{\partial z}\bigg)}
  -\varepsilon_{www}
  + SPS^{www}, \label{eq_www_hoc}
\end{eqnarray}
\end{linenomath*}
where $b=\alpha_T\theta-\beta_s s$ is the buoyancy (with $\alpha_T$ and $\beta_s$ representing the thermal expansion and haline contraction coefficients, respectively) and primed terms denote plume-scale turbulent fluctuations from the spatio-temporal mean over the grid-cell volume and time step of the GCM. This is slightly different from standard Reynolds averaging techniques where primed terms represent the turbulent fluctuations at all scales. As a result, these plume-scale budgets also contain $SPS$ terms which represent the effects of sub-plume scale motion on the  plume-scale properties. The sub-plume scale turbulence is assumed to be isotropic and is evolved separately (see Appendix \ref{sec:sps}).

These budgets contain turbulent transport terms which depend on higher-order turbulent moments ($\overline{w'^2\theta'}$, $\overline{w'^2s'}$ and $\overline{w'^4}$) as well as other moments ($\overline{\theta'^2}$, $\overline{s'^2}$, $\overline{\theta's'}$) that must be diagnosed in order to close the equation set. We will discuss how the ADC scheme estimates these moments in the remainder of this section. The budgets also contain production terms resulting from buoyancy effects and from turbulent mixing in the presence of vertical gradients, in addition to turbulent pressure ($p'$) terms and dissipative terms ($\varepsilon_{w\theta}$, $\varepsilon_{ws}$, $\varepsilon_{ww}$ and $\varepsilon_{www}$). The pressure and dissipative terms must also be parameterized to close this equation set, and these parameterizations are discussed in Appendices \ref{sec:pressure} and \ref{sec:sps} respectively.

\subsection{Physical basis of the Assumed Distribution Closure (ADC) scheme}

The ADC scheme combines ideas from MFC and HOC schemes. Mass-flux closures were originally developed to parameterize atmospheric convection and they typically distinguish between fluid that is in upwelling plumes (or updrafts) and the surrounding fluid in the environment or in downwelling plumes \cite{Arakawa1969, Ooyama1971}. Each type of plume covers a specific fraction of the total area and has a specific vertical profile of thermodynamic and dynamic variables. These properties vary with vertical position and can evolve in time, and inter-plume property differences are used to estimate vertical fluxes. In contrast, higher-order closure schemes evolve prognostic equations for turbulent statistics of different orders, including the vertical fluxes. Unfortunately, HOC schemes are computationally more expensive than MFC schemes due to the large number of equations that must be solved, and they require closure assumptions since the full equation set is infinite.

\citet{Lappen2001a} proposed the ADC method, where a small number of HOC equations are evolved to provide turbulent statistics that inform a MFC scheme, and then the MFC scheme is used to calculate higher-order statistics that close the HOC equation set. This method maintains the physical tractability of HOC and uses MFC to reduce the equation set, making the scheme more computationally efficient than standard HOC methods. The key element of the ADC scheme is that a distribution of properties must be assumed within each of the plume types to merge the MFC and HOC methods. Following \citet{Lappen2001a}, we assume that properties have a top-hat (or double-$\delta$) distribution where at a given height and time each variable can take one of two values, depending on whether that fluid is in an upwelling or a downwelling plume, and that within each plume all properties are horizontally uniform. In the next section we discuss how this assumption allows the high-order terms that appear in flux budgets such as Eqs. \ref{eq_wtheta_hoc}-\ref{eq_www_hoc} to be estimated, closing the budgets.

\begin{figure}[h!]
\centering
\includegraphics[width=0.38\textwidth]{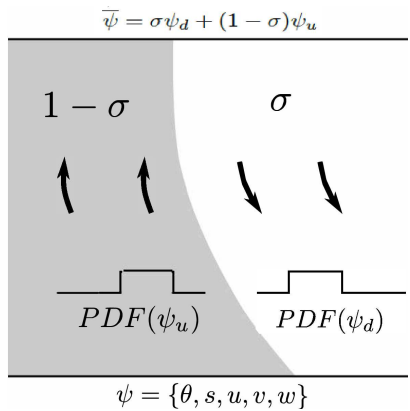}
\caption{Theoretical classification of plumes in a column in a mass-flux closure and the carried probability distributions associated with the plumes.}
\label{fig:massflux}
\end{figure}

\begin{figure}[h!]
\centering
\includegraphics[width=400pt]{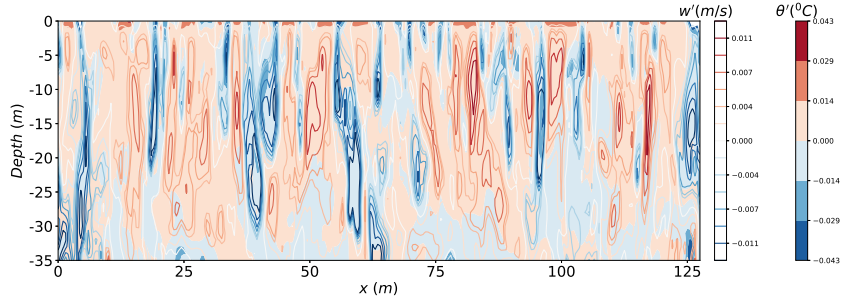}
\caption{Overlaid contours of instantaneous vertical velocity fluctuations $w'$ (lines) and temperature fluctuation $\theta'$ (filled) of an instantaneous vertical cross-section (y=64m) from three-dimensional LES fields. }
\label{fig:contour_xz}
\end{figure}

\begin{figure}[h!]
\centering
\includegraphics[width=0.8\textwidth]{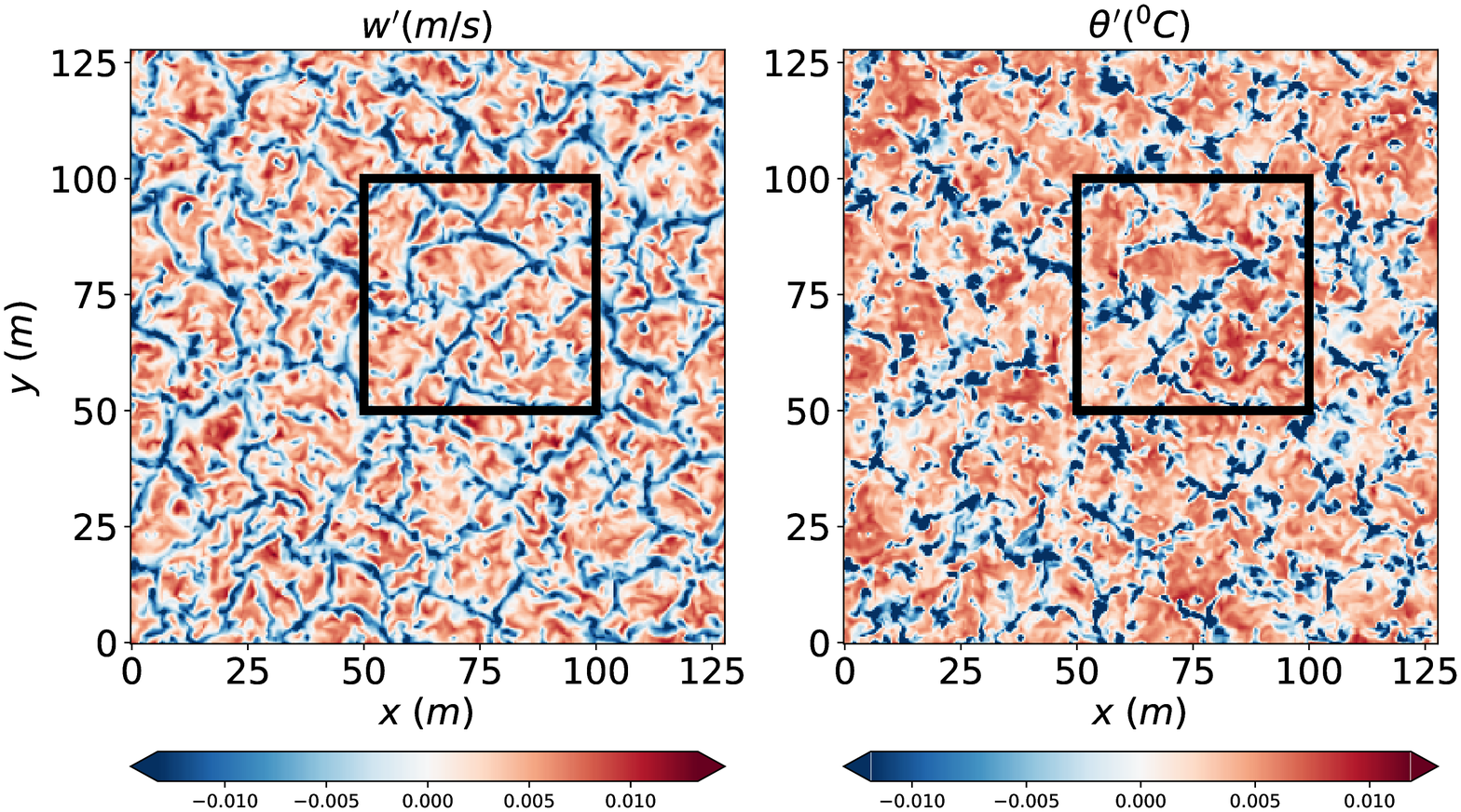}
\includegraphics[width=0.7\textwidth]{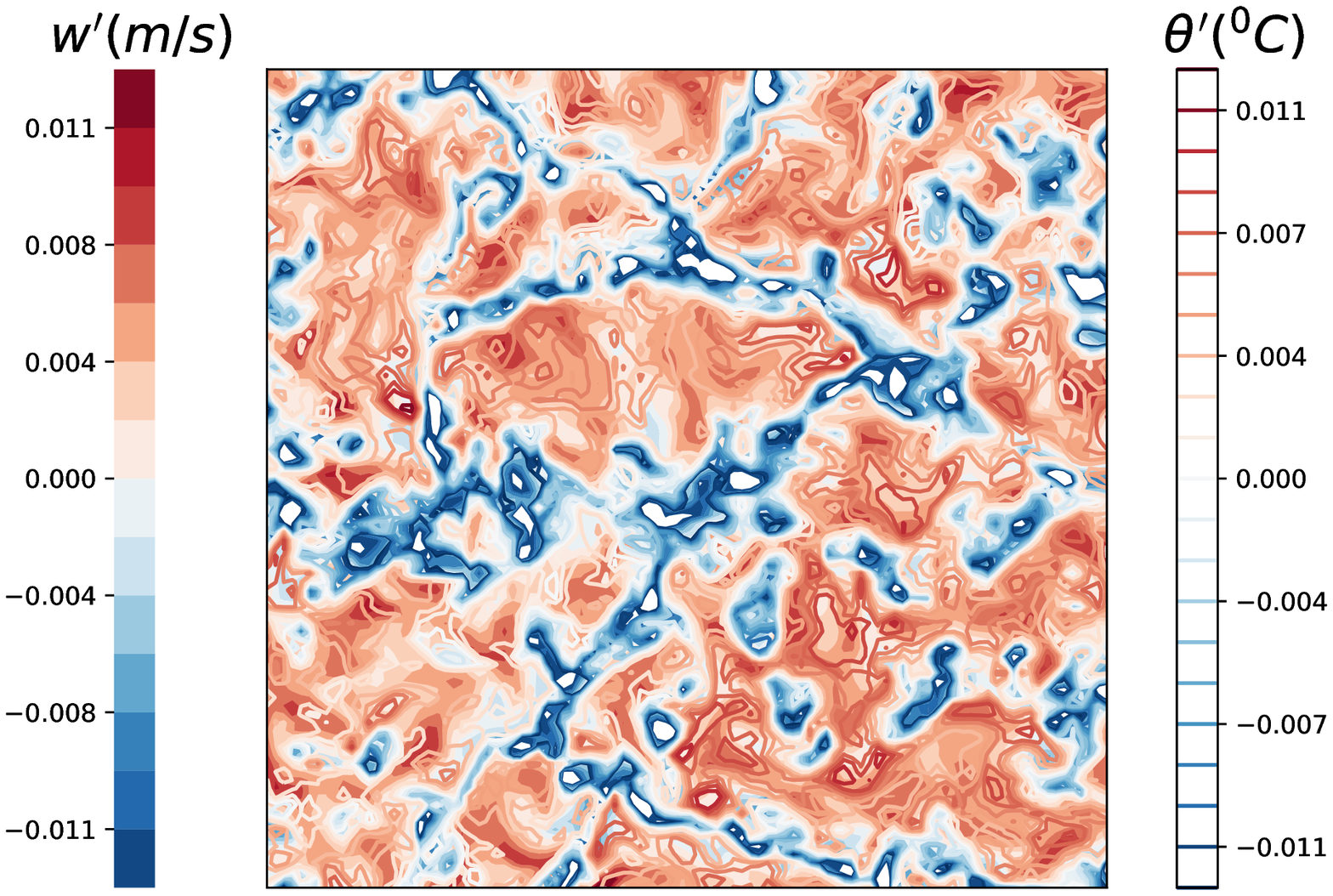}
\caption{ Top: instantaneous  velocity fluctuations ($w'$, left panel) and temperature fluctuations ($\theta'$, right panel) of a horizontal cross-section (z=-5m) from three-dimensional LES field and bottom: overlaid contour of instantaneous $w'$ (lines) and instantaneous $\theta'$ (filled) for the same horizontal cross-section as in top panel showing a strong co-relation. A zoomed section of top panel (marked by black box) is shown in the bottom panel for better visibility. }
\label{fig:contour_xy}
\end{figure}

The ADC method has not been previously applied to oceanic turbulence, but plumes are also common in the upper ocean under various dynamical regimes, including oceanic convection which is the focus of the present study (we discuss future application of the ADC scheme to other ocean regimes in section \ref{sec:diss_n_concl_lang}). Figure \ref{fig:massflux} shows key features of the new ocean ADC scheme, including the assumption that all the fluid within a grid cell is split between two types of plume (upwelling or downwelling) that cover different area fractions of that cell ($1-\sigma$ and $\sigma$ respectively), and that each of these plume types has specific properties which vary with depth and in time (as shown in Figure \ref{fig:massflux}). The relative area and vertical speed of each plume type compensates to ensure mass continuity in each grid cell. This two-plume physical model (or the top-hat PDF assumption) is supported by LES of convective turbulence in the ocean, which show strong correlations between temperature and the sign of vertical motion in both vertical cross-sections (Figure \ref{fig:contour_xz}) and horizontal cross-sections (Figure \ref{fig:contour_xy}). The present study will demonstrate the utility of a two-plume MFC scheme for representing oceanic convection, but it should also be noted that more complex atmospheric MFC schemes exist which have several plume types or include non-uniform property distributions within each plume, which could provide future development directions for this oceanic scheme (section \ref{sec:diss_n_concl_pdf}).

\subsection{Estimating higher-order-moments with the ADC scheme}

Turbulence statistics require information about fluctuations of properties from their background mean value ($\overline{\psi}$). In the two-plume ADC formulation where we know the properties in downwelling and upwelling plumes ($\psi_d$ and $\psi_u$ respectively), the background mean and turbulent fluctuations can be constructed from the plume properties and areas,
\begin{linenomath*}
\begin{equation}
   \overline{\psi}=\sigma\psi_d+(1-\sigma)\psi_u,
   \end{equation}
   \begin{equation}
    \psi'_d=\psi_d-\overline{\psi}, \qquad \textrm{and} \qquad \psi'_u=\psi_u-\overline{\psi}.
\end{equation}
\end{linenomath*}

From this two-plume model, it is possible to calculate various turbulence statistics using the plume properties. For example, some of the turbulence statistics in the ADC prognostic equations (Eqs. \ref{eq_wtheta_hoc}-\ref{eq_www_hoc}) can be diagnosed as,
\begin{linenomath*}
\begin{eqnarray}
\overline{\theta'^2}&=& \sigma(\theta_d-\overline{\theta})^2+(1-\sigma)(\theta_u-\overline{\theta})^2 \label{eq_TT} \\
&=&\sigma(1-\sigma)(\theta_u-\theta_d)^2,
\label{eq:theta_var} \\
\overline{s'^2}&=& \sigma(1-\sigma)(s_u-s_d)^2,
\label{eq:s_var} \\
\overline{\theta's'}&=& \sigma(1-\sigma)(s_u-s_d)(\theta_u-\theta_d),
\label{eq:stheta} \\
\overline{w'^2s'} &=& - (1-2\sigma)M_c(w_u - w_d)(s_u - s_d) \label{eq_wws}, \\
\overline{w'^2\theta'} &=& - (1-2\sigma)M_c(w_u - w_d)(\theta_u - \theta_d) \label{eq_wwT},
\end{eqnarray}
\end{linenomath*}
where the convective mass flux $M_c=\sigma(1-\sigma)(w_u-w_d)$. Note that in this ocean scheme, the diagnostic relations for the third-order moments (Eqs. \ref{eq_wws}-\ref{eq_wwT}) contain a minus sign which is not present in atmospheric schemes such as \citet{Lappen2001a}. This is because the $\sigma$ in the ocean scheme is the area of downwelling plumes instead of upwelling plumes, a choice made to ensure $\sigma$ represents the intense convecting plumes in both systems.

Calculating the above turbulence statistics requires the downwelling plume area ($\sigma$), convective mass flux ($M_c$), and plume differences in temperature and salinity. These quantities are diagnosed from the turbulence statistics evolved by the ADC scheme ($\overline{w'\theta'}$, $\overline{w's'}$, $\overline{w'^2}$ and $\overline{w'^3}$) via their own relations to the plume properties. The downwelling area fraction and mass-flux are calculated as,
\begin{linenomath*}
\begin{equation}
\sigma=\frac{1}{2}\left(1+\frac{S_w}{\sqrt{4+S_w^2}}\right), \qquad M_c=\sigma(1-\sigma)(w_u-w_d)=\frac{\sqrt{\overline{w'^2}}}{\sqrt{4+S_w^2}},
\label{eq_sigma_Mc_diagnostic}
\end{equation}
\end{linenomath*}
where $S_w=\overline{w'^3}/(\overline{w'^2})^{3/2}$ is the skewness of the plume-scale turbulence. The skewness is typically negative in the upper ocean since downward motions are more intense than upward motions. A large negative skewness is a feature of oceanic convection and results in $\sigma \rightarrow 0$, because downwelling plumes are relatively strong and cover a smaller area fraction as a result of mass continuity. Under weak convection, the skewness approaches zero and $\sigma$ approaches $0.5$ as the third-order transport terms representing non-local mixing vanish leaving only local gradient terms in the variance equation \cite{Lappen2001a}. This is the limit of local (diffusive) mixing in the scheme. This equation form restricts $\sigma\in(0,1)$.
The tracer fluxes are then used to diagnose plume differences in temperature and salinity,
\begin{linenomath*}
\begin{eqnarray}
    (\theta_u-\theta_d)=\overline{w'\theta'}/M_c \quad \textrm{and} \quad (s_u-s_d)=\overline{w's'}/M_c.
\label{eq_psiud}
\end{eqnarray}
\end{linenomath*}

The MFC scheme can also directly diagnose fourth-order statistics, such as $\overline{w'^4}$ which appears in Eq. \ref{eq_www_hoc}. However, for this term we follow \citet{Gryanik2002} and use a slightly modified estimate for this term,
\begin{linenomath*}
\begin{equation}
\overline{w'^4}=M_c(1-\sigma+\sigma^2)(w_u-w_d)^3=\left(3+S_w^2\right)\big(\overline{w'^2}\big)^2.
\label{eq:w4_GH}
\end{equation}
\end{linenomath*}
The formulation given in equation (\ref{eq:w4_GH}) is essentially identical to the MFC form [$M_c(1-3\sigma+3\sigma^2)(w_d-w_u)^3)$] under highly-skewed (convective) conditions, but it transitions to local mixing as $S_w\rightarrow0$, where $\overline{w'^4}\approx 3 [\overline{w'^2}]^2$, following the quasi-normal approximation (QNA).

\begin{figure}[h!]
\centering
\includegraphics[width=1\textwidth]{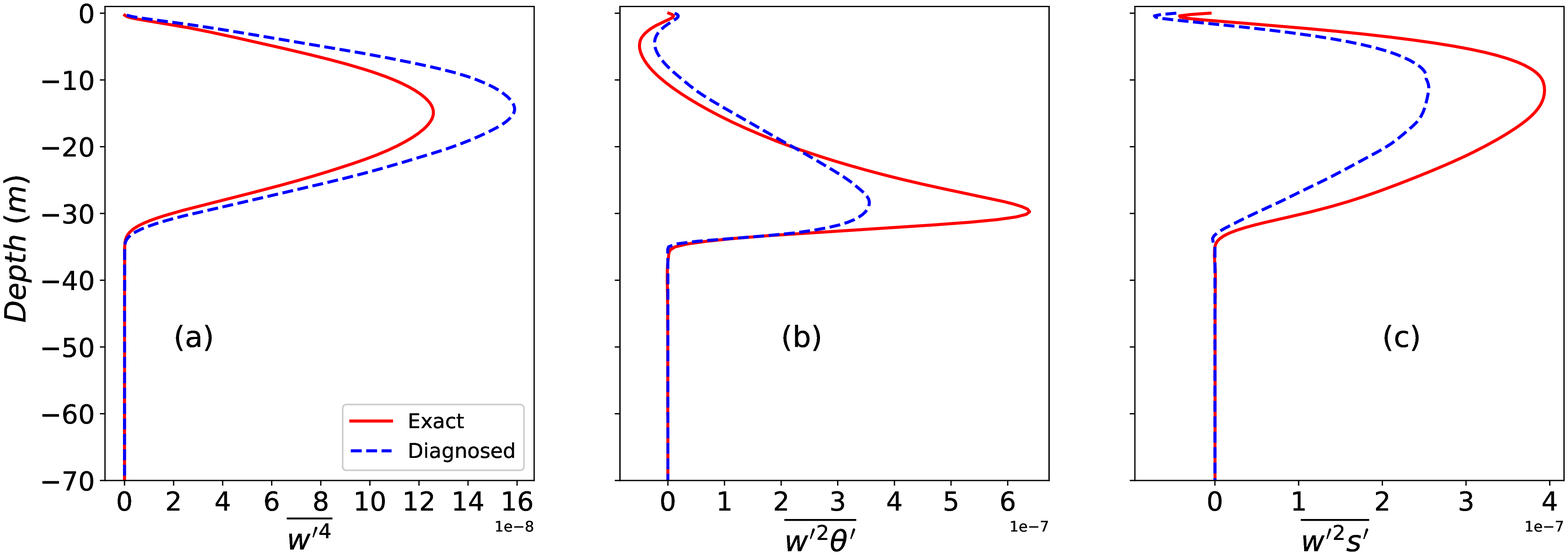}

\caption {A comparison of LES higher-order moments and their estimates diagnosed for a convective condition (test case T1S15)}. 
\label{fig:moment}
\end{figure}

In figure \ref{fig:moment} we compare the vertical profiles of $\overline{w'^4}$, $\overline{w'^2\theta'}$, and $\overline{w'^2s'}$ from an LES of convection (T1S15, Table \ref{tab:table1}) against the diagnostic relations of Eqs. \ref{eq_wws}, \ref{eq_wwT} and \ref{eq:w4_GH}. The mass-flux and plume properties were diagnosed from LES profiles of $\overline{w'^2}$, $\overline{w'^3}$, $\overline{w'\theta'}$ and $\overline{w's'}$ using Eqs. \ref{eq_sigma_Mc_diagnostic} and \ref{eq_psiud}. The ``exact'' profiles represent the direct output from LES and the ``diagnosed'' profiles are obtained through the above mentioned diagnostic relations from LES data. The good agreement between the diagnosed higher-order moments and exact higher-order moments (with a little discrepancy in the peak values) supports the use of the current scheme under convective conditions. The small discrepancy between these profiles are however acceptable for the accurate prediction of vertical turbulent fluxes of heat and salt (second-moment) and their mean states as shown in section \ref{sec:results}.

\subsection{Dissipation and sub-plume terms}

To parameterize the dissipative and sub-plume terms in the prognostic equations (Eqs. \ref{eq_wtheta_hoc}-\ref{eq_www_hoc}) we follow the methods used by \citet{Lappen2001a,Lappen2001c,Lappen2001b}. The dissipation of plume-scale properties is assumed to arise from lateral entrainment and detrainment, the mixing of fluid between the upwelling and downwelling plumes. This results in the following parameterizations,
\begin{linenomath*}
\begin{eqnarray}
 \varepsilon_{w\theta} &=& (E+D)(w_u-w_d)(\theta_u-\theta_d), \\
 \varepsilon_{ws} &=& (E+D)(w_u-w_d)(s_u-s_d), \\
 \varepsilon_{ww} &=& (E+D)(w_u-w_d)^2, \\
 \varepsilon_{www} &=& - \left[E(3\sigma-2)+D(3\sigma-1)\right](w_u-w_d)^3, \label{eq_dissipation_parameterization}
\end{eqnarray}
\end{linenomath*}
where $E$ and $D$ represent lateral entrainment and detrainment, respectively. The method for diagnosing $E$ and $D$ is based on turbulent mixing lengths and is described in Appendix \ref{sec_ED}. 

The sub-plume scale effects in Eqs. (\ref{eq_wtheta_hoc}-\ref{eq_www_hoc}) are parameterized as,
\begin{linenomath*}
\begin{eqnarray}
SPS^{w\theta}&=&\sigma(1-\sigma)(\theta_u - \theta_d)SPS_w - M_cSPS_{\theta}, \\
SPS^{ws}&=&\sigma(1-\sigma)(s_u - s_d)SPS_w - M_cSPS_{s}, \\
SPS^{ww}&=&2M_c SPS_w, \\
SPS^{www}&=& - 3(1-2\sigma)M_c(w_u-w_d)\:SPS_w,
 \label{eq_sps_parameterizations}
\end{eqnarray}
\end{linenomath*}
where $SPS_w$, $SPS_{\theta}$ and $SPS_s$ are terms that depend on sub-plume turbulent kinetic energy and fluxes. These terms are detailed in Appendix \ref{sec:sps}.

\subsection{Numerical implementation}
In summary, the new ADC scheme solves eight prognostic equations. These equations span four core variables (Eqs. \ref{eq_wtheta_hoc}-\ref{eq_www_hoc}), as well as the two horizontal TKE components that are required to parameterize the pressure terms (Eq. \ref{eq:ADC_uu_budget}) and the sub-plume scale TKE within each plume required to parameterize the dissipation terms (Eq. \ref{eq:sps_tke_tendency}).

The ADC scheme is implemented in the following progression:
\begin{enumerate}
    \item Apply boundary conditions and update plume properties ($\sigma$, $M_c$, $[\theta_u-\theta_d]$, $E$ etc.) using plume-scale statistics from the previous time step in Eqs. \ref{eq_sigma_Mc_diagnostic}-\ref{eq_psiud}.
    \item Calculate tendency terms of plume-scale and sub-plume-scale statistics (Eqs. \ref{eq_wtheta_hoc}-\ref{eq_www_hoc}, \ref{eq:ADC_uu_budget} and \ref{eq:sps_tke_tendency}) using the closure assumptions described above.
    \item Update plume-scale and sub-plume-scale statistics using their previous value and their current and prior tendency terms.
    \item Calculate total tracer fluxes by combining the plume-scale fluxes and sub-plume scale fluxes (Eq. \ref{eq_sps_flux}): $F_{\psi}=\overline{w'\psi'} + \sigma\overline{w'\psi'}_{sps, d}+(1-\sigma)\overline{w'\psi'}_{sps, u}$.
    \item Update mean profiles ($\overline{\theta}$ and $\overline{s}$) by applying the total fluxes to Eq. \ref{eq:GCM_equation}.
\end{enumerate}
The scheme is stepped forward in time by applying a third-order Adams-Bashforth (AB-3) method for turbulent statistics and a forward Euler method for the mean temperature and salinity profiles. The governing equations are solved on a vertically-staggered grid, where tracers are at the cell center, while second-order moments and vertical velocity are at the cell interface. The third- and fourth-order moments are defined at cell centers and cell interfaces, respectively. This facilitates collocated gradient calculations for higher-order moments.

As discussed here, we have presented the oceanic ADC scheme with significant modifications to the atmospheric ADC scheme to study OSBL turbulence. In our ADC scheme, temperature and salinity are evolved and their impact on density is accounted via the equation of state. The latter affects the buoyancy terms and the turbulent length scale calculation in the ADC scheme. This is in contrast to the atmospheric scheme where mean potential temperature and water vapor mixing ratio are the evolved tracers. We have also applied modified closure assumption for $\overline{w'^4}$ to create a smoother transition between non-local to local mixing (Eq. \ref{eq:w4_GH}). Buoyancy induced pressure fluctuations are accounted in the parameterization of pressure terms (Appendix \ref{sec:pressure}). In the next section we test the ADC scheme under oceanic convective conditions.

\section{Single column model test cases}
\label{sec:testcases}
For this study, the ADC model is implemented within the Model for Prediction Across Scales-Ocean  \cite<MPAS-Ocean,>[]{Ringler2013,Petersen2018}, the ocean component of the U.S. Department of Energy's Energy Exascale Earth System Model \cite<E3SM,>[]{Golaz2019}. We used a single column model formulation for testing. Test cases are initialized with a linear background stratification, either by a gradient of potential temperature ($T_z$), salinity $(S_z)$, or both. The initial stratification is defined using the squared buoyancy frequency, $N^2= -\frac{g}{\rho_0}\frac{\partial \overline{\rho}(z)}{\partial z}$, where $\rho(z)$ is density computed from a linear equation of state as follows
\begin{linenomath*}
\begin{equation}
    \label{eq:eos}
    \overline{\rho} = \rho_0(1-\alpha_T(\overline{\theta}-\theta_0)+\beta_S(\overline{S}-S_0)),   
\end{equation}
\end{linenomath*}
where $\alpha_T=2\times10^{-4}$~$^\circ$C$^{-1}$ and $\beta_S=8\times10^{-4}$~psu$^{-1}$ are the thermal expansion and haline contraction coefficients, respectively, $\rho_0=1026$~kg~m$^{-3}$ is the reference density for sea water, and $\theta_0=20$~$^\circ$C and $S_0=35$~psu are the reference temperature and salinity, respectively. 
Surface forcing is applied through a destabilizing heat flux ($Q_H)$ and/or a salinity flux $(Q_s$). To test the adequacy of our proposed ADC scheme for convective mixing, we have considered a number of different convective test cases with varying surface forcing and background stratification as detailed in Table \ref {tab:table1}. The test cases C1, C2, C4, and C16 represent free convective deepening of the mixed layer due to surface cooling. E1 and E4 are free convective test cases due to surface evaporation. Cases T1S0, T1S1, T1S3, and T1S15 show a combination of both surface cooling and evaporation. Test cases S1, S10, and S20 represent the impact of surface cooling on different degrees of stably stratified temperature profiles. We have used a uniform vertical grid spacing of 1~m over a 100~m depth with a turbulent time step of 1~s, except for T1S15 and  C16, where the turbulent time step is 0.5~s. Each simulation was run for a total of four days and the vertical profiles discussed in section \ref{sec:results} are time averages of the last six hours of the simulations, unless otherwise noted. The Deardroff convective velocity is defined as $w*=(\overline{w'b'}_0 h)^{1/3}$, where $\overline{w'b'}_0$ is the surface buoyancy flux and $h$ is the OSBL depth corresponding to final day of simulation for each test cases.

\begin{table}[h!]
\begin{center}
\caption{Summary of the test cases showing surface forcing and background stratification}
\label{tab:table1}
\begin{tabular}{c c c c c c c c}
\hline
 \text{Test name} & \text{Heat Flux}	& \text{Salinity Flux} & \text{$T_z$}	&\textbf{$S_z$}  & \textbf{$N^2$} & \textbf{$w^*$} &\\
 & $Q_H$ [W~m$^{-2}$] & $Q_S$ [kg~m$^{-2}$~s$^{-1}$]&  $[^\circ$C~m$^{-1}] $ & [psu~m$^{-1}$]  	& [s$^{-2}$] & [m~$s^{-1}$]\\
 \hline 
      	C1 		    & -50 			        & 0.0				        					& 0.1				& 0.0     & 1.96e-4   & 6.49e-3\\
      	C2 		    & -100       			& 0.0				       				    	& 0.1				& 0.0     & 1.96e-4 & 9.20e-3\\
      	C4 		    & -200 			        & 0.0			        						& 0.1				& 0.0     & 1.96e-4 & 1.30e-2\\
      	C16 	        & -800 			        & 0.0			        						& 0.1				& 0.0     & 1.96e-4 & 2.60e-2\\
      
        E1 		       & 0.0 			        & 8.9e-5			        					& 0.0				& -0.025  & 1.96e-4 & 6.37e-3\\
        E4 		       & 0.0       			    & 3.5e-4 		    	  				    	& 0.0				& -0.025  & 1.96e-4 & 1.28e-2\\
      	  
       	S1 	           & -100 			        & 0.0				         					& 0.01				& 0.0  &  1.96e-5 & 1.35e-2\\
      	S10 		   & -100       		    & 0.0				       				    	& 0.1				& 0.0  &  1.96e-4 & 9.20e-3\\
      	S20 		   & -100       			& 0.0				        					& 0.2				& 0.0  &  3.92e-4 &8.18e-3\\
      	 	  
        T1S0 	          & -50 			        & 0.0				                					& 0.05				& -0.025  & 2.94e-4 & 6.11e-3\\
      	T1S1 	          & -50       			    & 8.9e-5				         					& 0.05				& -0.025  & 2.94e-4 & 8.51e-3\\
      	T1S3 		      & -50			            & 2.6e-4 			        						& 0.05				& -0.025  & 2.94e-4 &1.19e-2\\
      	T1S15 	          & -50			            & 1.3e-3			         					& 0.05				& -0.025  & 2.94e-4 &2.38e-2\\
      	  \hline   
    \end{tabular}
  \end{center}
\end{table}

To verify the fidelity of our ADC scheme, we compare our results against three dimensional LES. We use the National Center for Atmospheric Research (NCAR) LES model \cite{McWilliams1997} that has been extensively used for simulating oceanic flows \cite{Sullivan2007,VanRoekel2012, vanRoekel2018,Hamlington2014,Smith2016,Smith2018}. Horizontal spatial derivatives are calculated spectrally, second-order finite differences are used for vertical velocity derivatives, third-order finite differences are used for vertical tracer derivatives, and a third-order Runge-Kutta time stepping is used with a constant Courant number. The sub-grid scale model is detailed in \citet{Sullivan2007}. The domain size is $128 \times 128 \times 128$~m, with a computation grid of size of $256^3$. Uniform grid spacing is used in the horizontal, while a stretched grid, starting with a surface layer thickness of $0.2$~m (with a gradual increase of grid spacing to $1$~m over $128$~m depth) is used in the vertical. Initial conditions and surface forcing are horizontally uniform and correspond to the same linear background stratification and fluxes as the ADC simulations (Table \ref{tab:table1}). Homogeneous Neumann boundary conditions are applied at the bottom of the LES domain. Similar to the ADC model results, the LES was run for a total of four days and vertical profiles discussed in section \ref{sec:results} are both horizontally averaged and time averaged over the last six hours of each simulation, unless otherwise noted.   

\section{Results }
\label{sec:results}
In this section, we present results from the ADC model and compare them to the LES data. We begin by discussing mean, plume, and second-order moment profiles. Next, we compare the ADC scheme with two widely used vertical mixing parameterizations in GCMs: KPP \cite{Large1994} and $k-\varepsilon$ \cite{Rodi1987}. Sensitivity of the ADC scheme to vertical resolution and time step is discussed in Appendix \ref{sec:reso_sensitivity}. 
\begin{figure}[h]
\centering
\includegraphics[width=13cm]{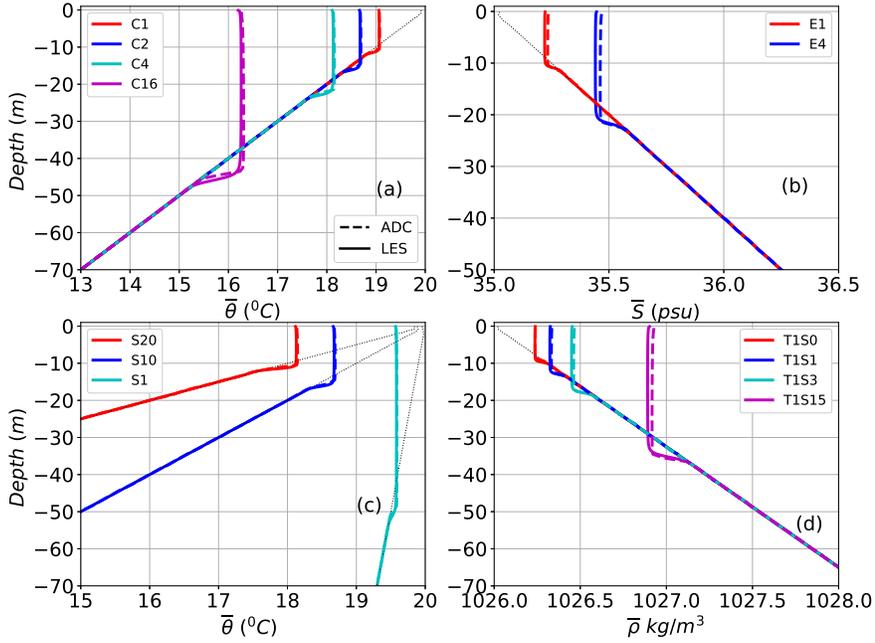}
\caption{Mean vertical profiles of (a) temperature for surface heat flux cases (C1, C2, C4, and C16), (b) salinity for surface salinity flux cases (E1 and E4), (c) temperature for initial stratification cases (S1, S10, and S20), and (d) density for combined surface heat flux and salinity flux cases (T1S0, T1S1, T1S3, and T1S15) for both LES (solid lines) and the ADC model (dashed lines) on the final day of the simulations. Initial profiles are shown as a black dotted line and colors indicate different test cases.}
\label{fig:mean_profiles_1}
\end{figure}
\subsection{Mean, plume, and second-order moments}
Figure \ref{fig:mean_profiles_1} shows mean profiles of temperature, salinity, and density for both the ADC model and corresponding LES for all test cases in Table \ref{tab:table1}. Here, for cases with only a surface heat flux (C1, C2, C4, C16, S1, S10, and S20), we show mean temperature profiles, and, for cases with only a surface salinity flux (E1 and E4), we show mean salinity profiles. For test cases with combined effects of surface heat and salinity flux (T1S0, T1S1, T1S3, and T1S15), density profiles are shown. Deepening of the boundary layer for a free convective mixing case is a function of surface fluxes and background stratification \cite{Mironov1990,Burchard2001}. As seen in Figure \ref{fig:mean_profiles_1}, stronger forcing and weaker stratification both result in progressively deeper mixed layers. The ADC model (dashed lines) successfully captures the growth of the mixed layer depth and the characteristics of the entrainment layer seen in LES (solid lines), across all forcing and initial condition scenarios.
\begin{figure}[h!]
\centering
\relax
\includegraphics[width=13cm]{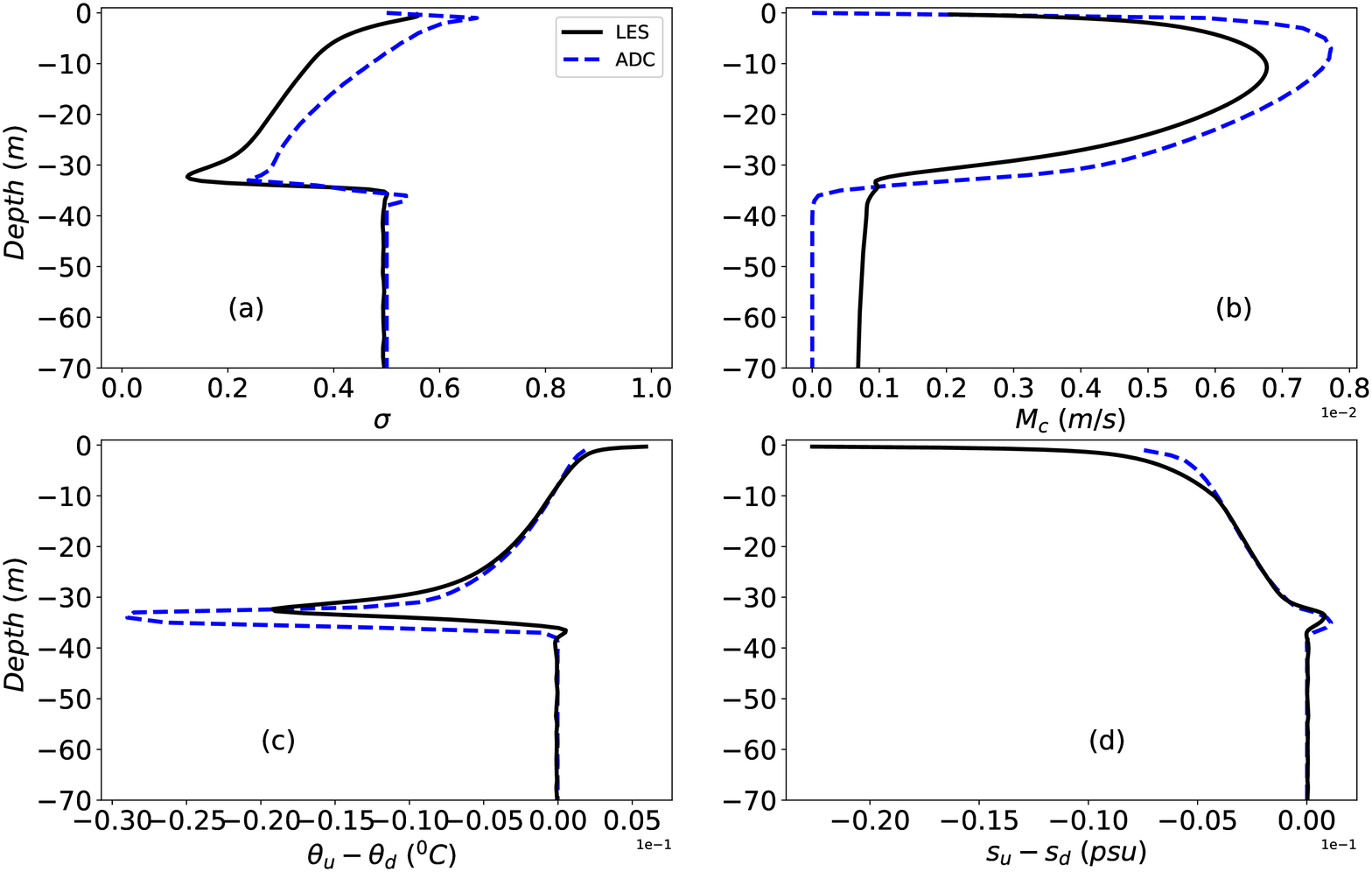}
\caption{Mean vertical profiles of (a) area fraction, (b) convective mass flux, (c) temperature upwelling-downwelling difference, and (d) salinity upwelling-downwelling difference for the T1S15 test case for both LES (black solid line) and the ADC model (blue dashed line) on the final day of the simulations. }
\label{fig:plume1}
\end{figure}
 
In Figure \ref{fig:plume1}, we show the plume profiles $\sigma$, $M_c$, $(\theta_u-\theta_d)$, and $(s_u-s_d)$ for both the ADC model and LES for a sample test case T1S15. As discussed in section \ref{sec:methods}, the higher-order moments are diagnostically determined through the ADC scheme if these plume properties are known. For the purpose of comparison, the LES plume profiles are obtained from the velocity skewness and turbulent flux statistics derived using profiles of $\overline{w'^2}$, $\overline{w'^3}$, $\overline{w'\theta'}$, and $\overline{w's'}$ (equations (\ref{eq_sigma_Mc_diagnostic}) and (\ref{eq_psiud})), while the ADC profiles are the ADC model output. All of the profiles are averaged over the last twelve hours. ADC profiles for $\sigma$, $M_c$, $(\theta_u-\theta_d)$, and $(s_u-s_d)$ are comparable to LES with small discrepancies. It should be noted that $M_c$ is non-zero in the interior for LES data. $M_c$ is obtained from the velocity variance $\overline{w'^2}$ profiles which are non-zero in the interior for the LES data (see figure \ref{fig:w2_profiles}) due to the presence of internal-wave induced small vertical velocities. The convective mass flux has a roughly parabolic profile and $M_c$ is maximum slightly above the mid-level of the boundary layer, with much smaller values near the surface and the entrainment layer. As shown in both the LES and ADC, $\sigma<0.5$ for the bulk of the boundary layer, corresponding to a large negative skewness (strong downwellings) and, by definition, $\sigma=0.5$ at the surface and below the mixed layer where skewness is zero (see equation (\ref{eq_sigma_Mc_diagnostic})). For a closure to capture these regime changes, it must adaptively reduce to a local diffusive type closure below the boundary layer as convective motion ceases, which the ADC model does well. The ADC scheme becomes traditional HOC with down-gradient moments when skewness is zero, as the third-moments representing non-local transport vanish for $\sigma=0.5$.  Similar results to those in Figure \ref{fig:plume1} are seen for all test cases (not shown).

\begin{figure}[h]
\centering
\includegraphics[width=13cm]{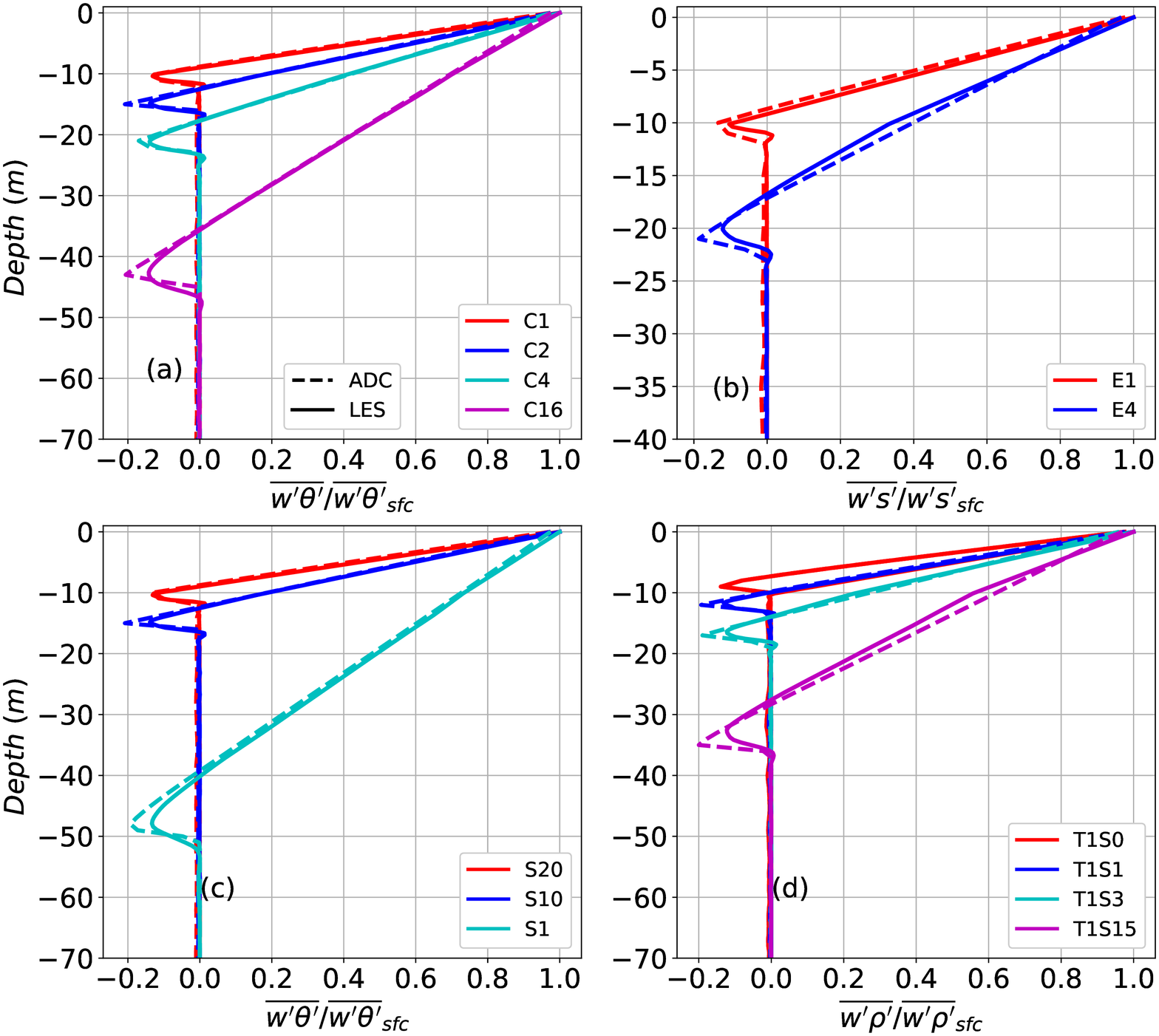}
\caption{Normalized vertical turbulent flux profiles of (a) heat for surface heat flux cases (C1,C2, C4, and C16), (b) salinity for surface salinity flux cases (E1 and E4), (c) heat for initial stratification cases (S1, S10, and S20), and (d) buoyancy for combined surface heat flux and salinity flux cases (T1S0, T1S1, T1S3, and T1S15) for both LES (solid lines) and the ADC model (dashed lines) on the final day of the simulations.}
\label{fig:wt}
\end{figure}

\begin{figure}[h]
\centering
\includegraphics[width=13cm]{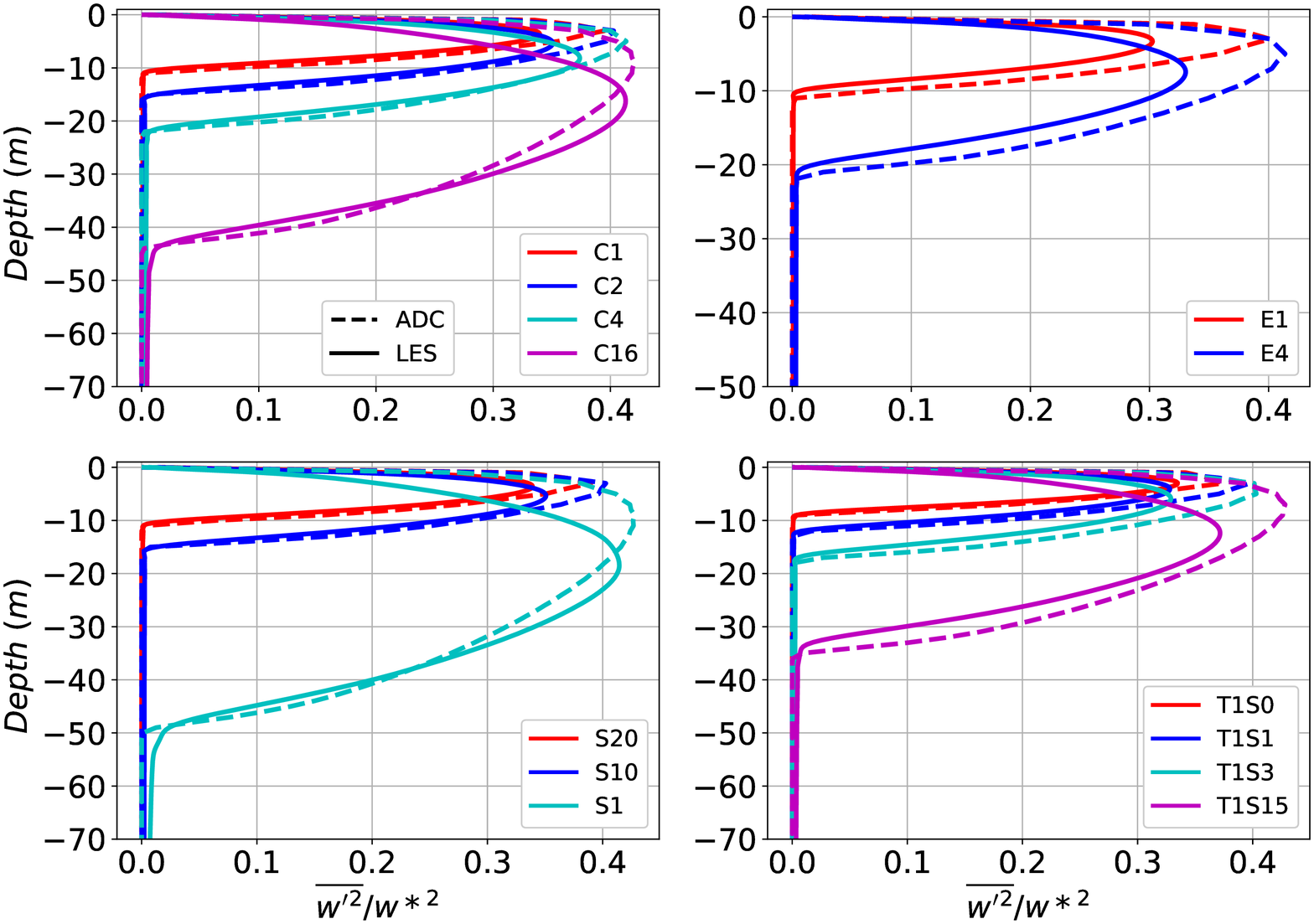}
\caption {Same as Figure \ref{fig:wt} but for vertical velocity variance normalized by Deardorff's convective velocity} 
\label{fig:w2_profiles}
\end{figure}

The turbulent heat-flux $(\overline{w'\theta'})$, salinity-flux $(\overline{w's'})$, and density-flux $(\overline{w'\rho'})$ for each test case with colors and layout analogous to Figure \ref{fig:mean_profiles_1} are shown in Figure \ref{fig:wt}, normalized by their respective surface fluxes. The ADC results show that the ratio of minimum turbulent flux in the entrainment layer to the surface flux is approximately $-0.2$ for all the test cases, which is typical for a convective boundary layer \cite{Large1994, vanRoekel2018}. Again, the excellent agreement between ADC and LES turbulent flux profiles suggests that the ADC is able to accurately simulate the height and thickness of the entrainment layer and hence the evolution of the OSBL. The two equation $k-\epsilon$ closure model often underestimates both height and thickness of this entrainment layer resulting in shallow bias in OSBL depth \cite{Burchard2001}.

Figure \ref{fig:w2_profiles} shows the vertical velocity variance $\overline{w'^2}$ normalized by Deardorff's convective velocity for all the test cases (same organization by case type as Figures \ref{fig:mean_profiles_1} and \ref{fig:wt}). The vertical velocity variance, constrained to zero at the surface, reaches its maximum within the boundary layer before decaying to zero below the boundary layer. Again, the ADC scheme, in comparison to the LES results, does a good job of capturing the strength and relative shape of $\overline{w'^2}$ for all the test cases. The only discrepancy is the small deviation in the depth location of the maximum for each case. For all the test cases, maximum $\overline{w'^2}$ corresponds to a depth $z/h \approx 0.3-0.36$ for LES results and $z/h\approx 0.2-0.3$ for ADC results, where $z$ is the depth of maximum velocity variance and $h$ is the OSBL depth diagnosed as the maximum in $N^2$. The ADC model results show a shallower and slightly larger maximum than the LES in most cases. Other studies have shown a maximum velocity variance corresponds to $z/h\approx 0.4$ for a convective boundary layer \cite{Lappen2001a, Zhou2019}. We also note that, maximum velocity variance for KPP is at $z/h\approx 0.33$ \cite{Large1994, Burchard2001}. 

\subsection{Comparison with other parameterizations}
To further demonstrate the utility of the ADC model, we compare results to two widely used vertical mixing parameterizations in GCMs: KPP \citep{Large1994} and $k-\varepsilon$ \cite{Rodi1987}. The generalized ocean turbulence model \cite<GOTM,>[]{Umlauf2005} is used to run single-column simulations with both KPP and $k-\varepsilon$ for all test cases in Table~\ref{tab:table1}. For the KPP model, we use the default CVMix \cite{Griffies2015} settings within GOTM, where the critical bulk Richardson number is set to $0.3$. Similarly, default values within GOTM are used for the $k-\varepsilon$ model, with the stability functions of \citet{Schumann1995}. Two vertical resolutions, $1$~m and $10$~m, are used and simulations are carried out for the same four day period. 

\begin{figure}[h!]
\centering
\includegraphics[width=8cm]{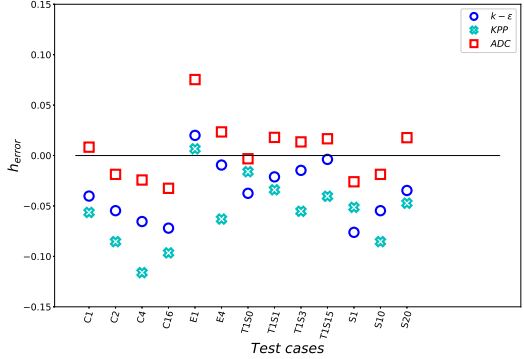}
\caption{Relative error in the OSBL depth (equation (\ref{eq:RE_osbl})) compared to LES results for the ADC (red squares), KPP (cyan x's), and $k-\varepsilon$ (blue circles) models averaged over the last twelve hours of each test case. }
\label{fig:osbl_rel_err}
\end{figure}

Figure \ref{fig:osbl_rel_err} shows the relative error in the OSBL depth compared to LES results for the ADC, KPP, and $k-\varepsilon$ models averaged over the last twelve hours of each test case (Table~\ref{tab:table1}) with a vertical resolution of $1$~m. The relative error in OSBL depth ($h$) is computed as
\begin{linenomath*}
\begin{eqnarray}
    \mathrm{h}_ \mathrm{error}=\frac{h_{model}-h_{LES}}{h_{LES}}. 
    \label{eq:RE_osbl}
\end{eqnarray}
\end{linenomath*}
\begin{figure}[h!]
\centering

\includegraphics[width=12cm]{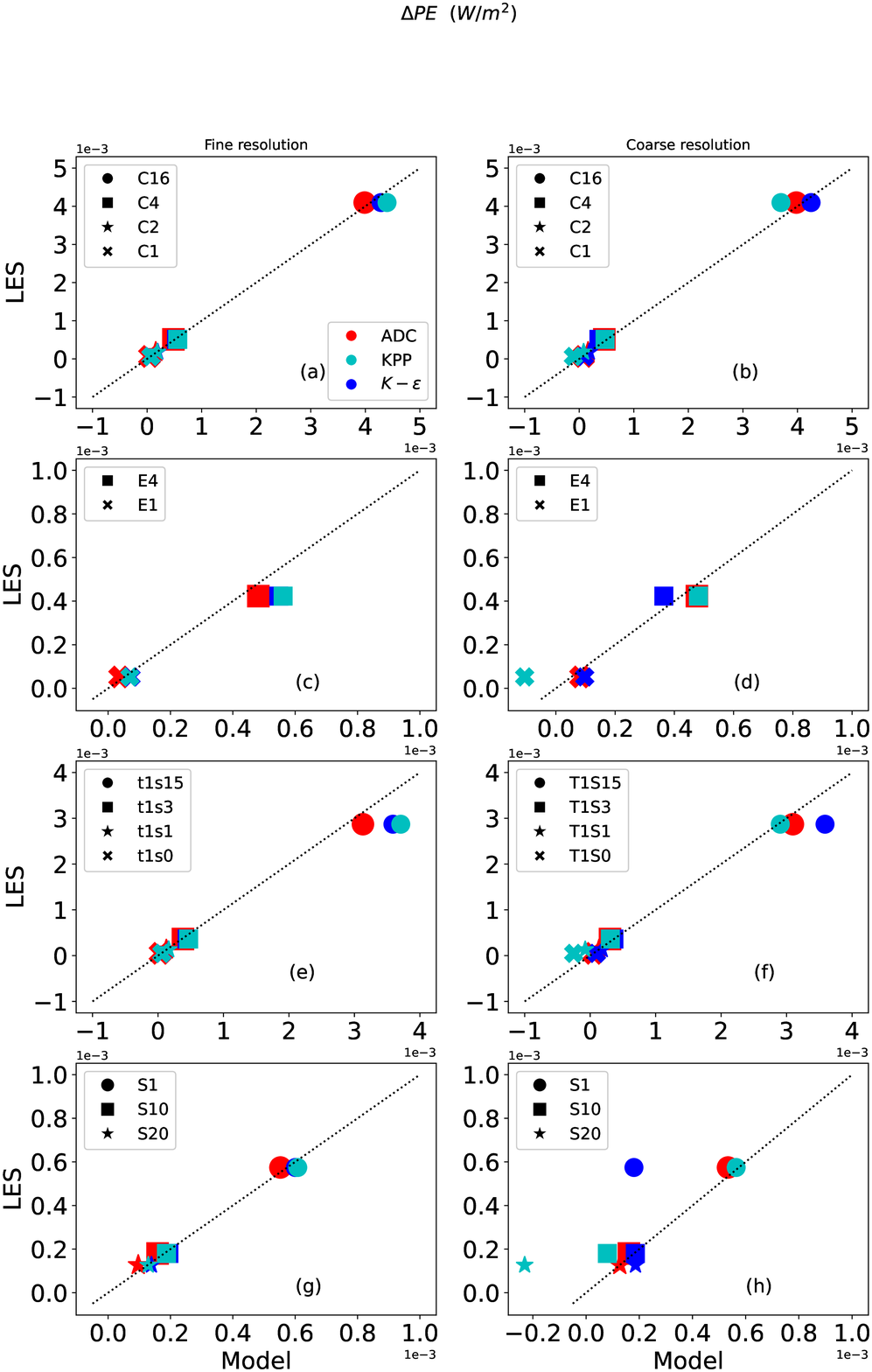}
\caption{Rate of change in depth-integrated potential energy ($W/m^2$) for LES (vertical axis) vs the ADC (red), $k-\varepsilon$ (blue), and KPP (cyan) models (horizontal axis) for the (a-b) surface heat flux, (c-d) surface salinity flux, (e-f) initial stratification, and (g-h) combined surface heat and salinity flux cases using (a, c, e, g) $1~m$ and (b, d, f, h) $10~m$ vertical resolutions. The dashed black line shows a one-to-one comparison of the LES and the closure models.}
\label{fig:pe_ke_kpp_adc}
\end{figure}
\begin{figure}[h!]
\centering

\includegraphics[width=15cm]{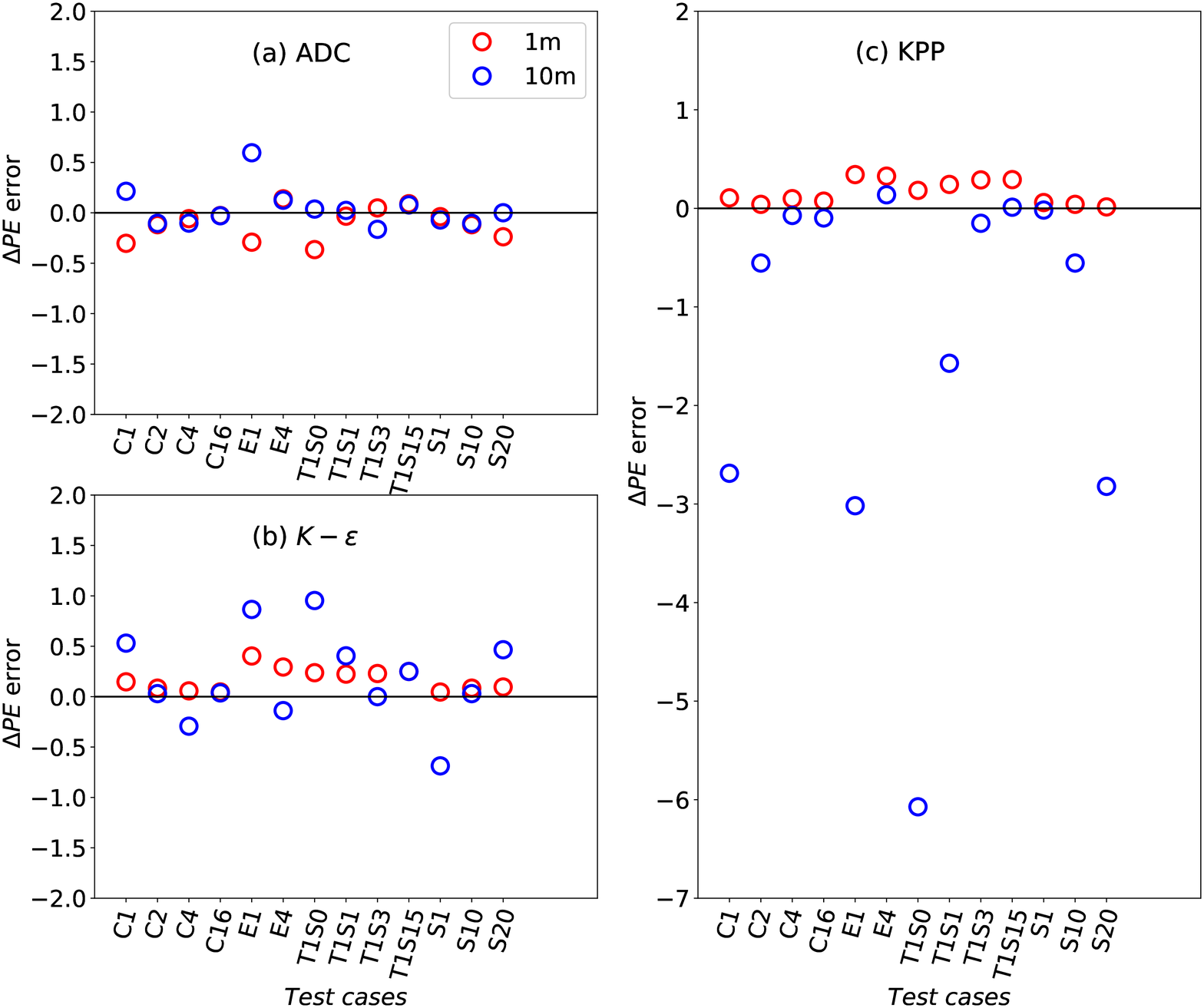}

\caption{Relative rate of change in depth-integrated potential energy for the (a) ADC , (b) $k-\varepsilon$, and (c) KPP models with vertical resolutions of $1$~m (red circles) and $10$~m (blue circles).} 
\label{fig:re_pe_all}
\end{figure}
The subscript $model$ refers to the respective ADC, KPP, and $k-\varepsilon$ parameterization schemes and the OSBL depth is defined as the depth of maximum $N^2$. Negative relative error corresponds to a shallow bias in the OSBL depth compared to LES. As shown in Figure \ref{fig:osbl_rel_err}, KPP (cyan x's) and $k-\varepsilon$ (blue circles) show a consistent shallow bias in OSBL depth (except for the E1 test case), where KPP shows a maximum error of $~11\%$ (for C4 case). In contrast, ADC results (red squares) have a small absolute relative error of $\leq 2.5\%$ across all test cases (except for the E1 test case). These errors are a mix of shallow and deep minor biases. The E1 test case, however, captures both mean and turbulent profiles within acceptable error as shown in Figures \ref{fig:mean_profiles_1} and \ref{fig:wt}, respectively. In general, the OSBL depth is sensitive to the method used to diagnose the stratification of the entrainment layer (or maximum $N^2$) \cite{vanRoekel2018, Reichl2019}. In KPP, the OSBL depth is the master parameter \cite{Large1994} and is subject to large biases at coarse resolutions depending on the method of interpolation and number of grid points in the boundary layer. 

To better compare the three mixing schemes across changes in vertical resolutions and the energetics of turbulent mixing that modulates the growth of the OSBL depth  we compute the change in integrated potential energy \cite{Reichl2018, Reichl2019} over the duration of the simulation. The rate of change in the depth integrated potential energy from its initial profile for all the test cases and across all the resolutions is shown in Figure \ref{fig:pe_change_adc_les} to investigate the effects of resolution and surface forcing on the energetic constraint of the OSBL. The rate of change in the depth integrated potential energy is defined as
\begin{linenomath*}
\begin{eqnarray}
 \Delta PE= \frac{1}{\Delta t}\int_{z1}^{z2} \left[\overline{\rho} (z)-\overline{\rho_{ini}} (z)\right]gz \,dz,
 \label{eq:deltape}
\end{eqnarray}
\end{linenomath*}
where $\overline{\rho_{ini}}(z)$ is the initial density profile, $\overline{\rho} (z)$ is the density at the end of the four day simulation averaged over the last twelve hours, and $z$ is the depth. Integration is done from $z1=0$~m to $z2=-100$~m. The change in potential energy quantifies mixing within the OSBL due to both surface fluxes and entrainment from below. A deeper mixed layer corresponds to a larger $\Delta PE$. The relative difference in the rate of change in integrated potential energy between the LES and ADC model is defined as 
\begin{linenomath*}
\begin{eqnarray}
  \Delta PE_ \mathrm{{error}}=\frac{\Delta PE_{model}-\Delta PE_{LES}}{\Delta PE_{LES}}.
 \label{eq:RE_deltaPE}
\end{eqnarray}
\end{linenomath*}

Figure \ref{fig:pe_ke_kpp_adc} shows the rate of change in the depth integrated potential energy (equation (\ref{eq:deltape})) for the ADC (red markers), KPP (cyan markers), and $k-\varepsilon$ (blue markers) models in comparison to the LES results for both a fine vertical resolution ($1$~m) and coarse vertical resolution ($10$~m). The potential energy is integrated from $100$~m to the surface and results are averaged over the last twelve hours. The ADC model compares well against LES results for both resolutions and across all the test cases, while KPP and $k-\varepsilon$ underestimate mixing at the higher vertical resolution. This is more evident for strong surface forcing or weaker background stratification where the OSBL depth is deeper. For the coarse vertical resolution, KPP and $k-\varepsilon$ are unable to accurately simulate turbulent mixing across various test cases, suggesting a very strong resolution dependence for both of these models. Figure \ref{fig:re_pe_all}  shows the relative $\Delta$ $PE$Error (equation (\ref{eq:RE_deltaPE})) corresponding to Figure \ref{fig:pe_ke_kpp_adc} for the (a) $k-\varepsilon$, (b) KPP, and (c) ADC  parameterization schemes across all the test cases and the two vertical resolutions ($1$~m and $10$~m). For most of the test cases (except for C1, E1, T1S0, S20), the relative error for the ADC scheme is less than $15\%$ for both fine and coarse resolutions. For the test cases C1, E1, T1S0, and S20, because the OSBL depth is only about $10$~m at the end of four days (Figure \ref{fig:mean_profiles_1}), there are not enough data points to make a fair comparison. Even accounting for these four test cases, the ADC scheme shows the least sensitivity to vertical resolution, while for both KPP and $k-\varepsilon$ the sensitivity to vertical resolution is case specific and the maximum error is much larger. The detailed study of insensitivity to vertical resolution of ADC scheme is discussed in appendix \ref{sec:reso_sensitivity}.

\section{Discussions}
\label{sec:diss_n_concl}
This paper presents a new, physically-motivated, ocean ADC scheme to parameterize vertical turbulent mixing in the upper ocean. The ADC scheme combines a MFC and HOC using an assumed (top-hat) distribution of properties within each type of plume (upwelling or downwelling). The unified framework of ADC is suitable for regimes where large-scale convection and small-scale turbulence co-exist, as is the case in the ocean surface boundary layer (OSBL). \citet{Lappen2001a, Lappen2001c, Lappen2001b} have successfully tested the ADC scheme for the atmospheric boundary layer. Here, we have extended the closure scheme to parameterize OSBL turbulence by making ocean-appropriate modifications and expanding closure assumptions. We validated this new scheme using high-resolution three-dimensional LES. The ADC model was shown to effectively simulate convective turbulence in the OSBL, reproducing the evolution of mean and turbulent flux profiles for a number of convective test cases, across different vertical resolutions and time steps.

We tested the viability of the ADC scheme for practical applications by conducting tests that varied the vertical grid spacing across four resolutions (1~m, 2~m, 5~m, 10~m) that are comparable to resolutions commonly used in regional and global ocean simulations. The time evolution of mean profiles for different strengths of surface forcing and background stratification showed excellent agreement with LES data. Many common mixing schemes show sensitivity to vertical resolution \cite{Li2019}, so we compared the resolution-sensitivity of the ADC scheme to that of the $k-\varepsilon$ \cite{Rodi1987} and KPP \cite{Large1994} mixing schemes by using a fine (1~m) and coarse (10~m) resolution (Figures \ref{fig:pe_ke_kpp_adc} and \ref{fig:re_pe_all}). Looking at the change in depth-integrated potential energy from the initial condition, we saw an insensitivity of the ADC scheme to vertical resolution, which is encouraging for use in GCMs. The rest of this section discusses the advantages and limitations of the ADC scheme along with potential future directions extension of the ocean ADC scheme.

\subsection{Advantages and limitations of ADC}
An advantage of ADC scheme over traditional HOC schemes is its requirement of fewer prognostic equations. All dynamic and thermodynamic quantities in the ADC scheme are represented with a top-hat PDF which allows the diagnosis of second-, third- and fourth-order moments that are consistent with one another and realizable (Eqs. \ref{eq_TT}-\ref{eq:w4_GH}). The prognostic equations used are term-by-term consistent with HOC/Reynolds-averaged equations \citep{Lappen2001a}. The higher-order moments that represent advective transport are diagnosed in a manner that inherently represents non-local convective transport (these terms are generally parameterized as locally diffusive in HOC). The small-scale mixing in the ADC scheme is represented through sub-plume scale parameterizations for turbulence that is not resolved explicitly by the mass-flux formulation. The ADC scheme is insensitive to changes in vertical resolutions and the turbulent time step. However, we note that the time steps used in this study are very small compared to that typically used in a GCM. The primary goal of this work was to provide a physically-based mixing scheme that can be used across varying resolutions. Although ADC is more expensive than a two-equation closure (e.g. \citet{MY1982}), it is less expensive than a full second-order model (e.g., \citet{Canuto2007}). In the future it may be possible to replace several of the predictive equations by diagnostic relations with appropriate assumptions and conditions, which would move the time-stepping towards a matrix solution. We could also implement some of the terms implicitly which would allow for longer time step. 

The ADC scheme uses sub-cycles within a large model/GCM time-step as discussed in \citet{Golaz2002}. While the present ADC scheme analyses were conducted on standard central processing units (CPUs), the ADC scheme's code design and it's implementation in MPAS mean that it can easily be run using a graphics processing unit (GPU) environment, enhancing the performance of the scheme. In benchmarking tests, we have seen a 75 times speed-up of ADC on GPU when compared to CPU.

\subsection{Application of ADC to other ocean mixing regimes}
\label{sec:diss_n_concl_lang}
The present work demonstrates that an ADC scheme can emulate convective ocean mixing driven by surface buoyancy fluxes, which is ubiquitous across the global ocean. There are several other processes not included in the present analysis that drive mixing in the upper ocean and are of global significance. These processes include mixing produced by surface waves (including Langmuir turbulence) or by surface winds. Although they are not the focus of the present analysis, here we discuss how the ADC scheme may be utilized or extended to these other regimes in future work.

Both wind- and wave-driven mixing introduce a preferred horizontal direction, resulting in horizontal currents, horizontal momentum fluxes ($\overline{u'w'}$ and $\overline{v'w'}$), and anisotropic turbulence in the upper ocean. To capture these regimes, the present ocean ADC scheme would need to be expanded to emulate $\overline{u'w'}$ and $\overline{v'w'}$, and the evolution of horizontal currents. The presence of these fluxes and shear in the horizontal currents leads to extra terms in the prognostic equations (Eqs. \ref{eq_wtheta_hoc}-\ref{eq_www_hoc}) which require a multitude of other second- and third-order turbulent statistics that must also be emulated (e.g., $\overline{u'\theta'}$, $\overline{u'v'}$, $\overline{u'w'w'}$). The ADC scheme's closure assumptions would also need to be expanded to include the effects of these new terms. The task of adding these horizontal momentum effects to the ADC scheme is challenging because horizontal momentum can behave differently to other thermodynamic variables. \citet{Lappen2001a} discuss this from the perspective of atmospheric convection, where horizontal velocity fluctuations typically peak at the interface between plumes, rather than within a specific type of plume like the other thermodynamic variables. Under these conditions $u'$ and $v'$ are not correlated with $w'$, $\theta'$ and $s'$, making it questionable to apply mass flux methods to diagnose the new terms ($\overline{u'\theta'}$, $\overline{u'v'}$, $\overline{u'w'w'}$ etc.) in the same manner as the ADC diagnoses other turbulent statistics (Eqs. \ref{eq_TT}-\ref{eq_wwT}). As a result, \citet{Lappen2001a} used prognostic equations to emulate the second-order moments that contained horizontal momentum, and a down-gradient approximation for the analogous third-order terms in convection.

Fortunately, the ocean may provide a unique environment that is particularly amenable to mass-flux representation of horizontal momentum terms. Specifically, Langmuir turbulence is one of the largest sources of upper ocean mixing globally \citep{Belcher2012} and typically consists of counter-rotating horizontal vortices called Langmuir cells that create surface convergence zones above strong downwelling jets \citep{McWilliams1997,polton2007}. These downwelling jets drive non-local vertical transport and occur beneath a strong along-cell surface jet. This suggests that mass flux methods could be an effective way to emulate the non-local transport resulting from Langmuir turbulence. Perhaps more intriguingly, the co-location of downwelling jets and maximal horizontal momentum suggests that a mass flux representation of the horizontal momentum terms might be a feasible approach to emulating Langmuir turbulence with an ADC scheme. This topic will be investigated in future work.  

\subsection{Comparison to other mass-flux type closures in the OSBL }

Recently, an Eddy-Diffusivity-Mass-Flux (EDMF) scheme was developed to study oceanic deep convection \cite{Giordani2020} based upon an earlier atmospheric scheme \cite{Siebesma1995,    Soares2004}. EDMF  uses a mass-flux closure to capture non-local mixing unified with a local K-theory type closure to capture local mixing. Both the ADC and EDMF scheme use mass-flux closure concept, however, each has its own merits and limitations. Here we provide a brief comparison of these two mixing schemes for future explorations. 

The EDMF scheme uses more a traditional mass-flux parameterization where the convective plumes are assumed to be distinguished from their environment. This suggests that the  plumes have much smaller fractional area ($\sigma<0.1$) and convective mass flux can be defined as $M_c=\sigma w_d$. This assumption is appropriate for deep convection. However, planetary boundary layer turbulence studies have shown that convective plumes and their environment are not always distinguishable, resulting in $\sigma$ being much higher \cite{Lappen2001a}. By using a two-plume framework (upwelling and downwelling plumes), the ADC scheme employs a more and the convective mass flux is defined with the weighted average of both upwelling and downwelling plume properties, $M_c=\sigma(1-\sigma)( w_u-w_d$). This representation is better suited when both upwelling and downwelling plumes are turbulent. 

Within the EDMF scheme, the turbulent fluxes are parameterized as the sum of two terms: down-gradient diffusion from the environment and mass-flux convection from the plume motions. As such, when $\sigma=0$ or $w_d=0$ the plumes disappear and the diffusion regime dominates. The third-order moment terms in the prognostic equations of the ADC scheme that represent the non-local transport (e.g., $\overline{w'^2\theta'}= (1-2\sigma)M_c(w_u - w_d)(\theta_u - \theta_d)=0$) vanish when $\sigma=0.5$ and the ADC scheme becomes a traditional HOC with down-gradient moments representing turbulent fluxes of the diffusion regime only. EDMF relies on parameterization of eddy diffusivity for the diffusion regime while in ADC, the prognostic equations of the turbulent fluxes are solved.

Both EDMF and ADC use evolution equations to estimate $\sigma$ and plume velocities. EDMF does this directly by solving evolution equations for $\sigma$ and $w_d$ (or $M_c$), while the ocean ADC scheme uses an assumed distribution method to estimate $\sigma$ and $M_c$ from from the prognostically-evolved second- and third-order moments of vertical velocity. It is possible to formulate an evolution equation of $\sigma$ in the ADC scheme \cite{Lappen2001a}, but we have not explored this in the present study. We note that lateral entrainment and detrainment directly influence convective fractional area and plume velocity in EDMF, while in ADC they act as dissipation terms in the prognostic equations of the turbulent statistics. 

Both the ADC and EDMF schemes are designed to capture both local and non-local mixing. The EDMF scheme has successfully emulated oceanic deep convection, while the new ADC scheme has been evaluated against shallow boundary layers typical of more general open-ocean conditions. As discussed in Section \ref{sec:diss_n_concl_lang}, we plan to expand this ADC scheme to emulate additional mixing processes in these shallow boundary layers. A future comparison between the ADC and EDMF schemes, and in-situ data, could provide an interesting exploration into the use of a mass-flux approach in various oceanic forcing scenario.

\subsection{Choice of PDF in ADC}
\label{sec:diss_n_concl_pdf}
The new ocean ADC scheme uses an assumed joint probability distribution for vertical velocity and thermodynamic variables. The equations that are typically used in the HOC models are then derived by integrating over the distribution with a mass-flux representation of higher-order moments. All of the parameters of the distribution are determined from the predicted moments; thereafter the joint distribution is effectively known, and so any and all moments can be constructed in a self-consistent manner \cite{Lappen2001a}. In this way, the scheme avoids the common closure problem caused by higher moments. As direct prediction of the actual sub-grid scale joint PDF is computationally expensive, one has to rely on an assumed PDF from a pre-selected family of PDFs \cite{Golaz2002}. Therefore, the success of an assumed distribution closure scheme depends on the choice of the assumed PDF. 

In the new ocean ADC scheme we have used a top-hat joint PDF (double-$\delta$) between vertical velocity and thermodynamic variables, which is supported by LES results showing their strong correlation in both upwelling and downwelling plumes. A two-plume mass-flux closure is also equivalent to a double-$\delta$ PDF representing the upwelling and downwelling plumes. In this way even though a double-$\delta$ PDF representation is the simplest out of all potential PDFs, it has the advantage of exactly closing the turbulent flux terms of HOC using the mass-flux framework. We note that, research in the atmosphere has begun to introduce complex PDFs such as double Gaussian \cite{Golaz2002,Larson2002, Fitch2019} and tri-variate Gaussian \cite{Firl2015}. Since ADC schemes have flexibility in their choice of PDFs without major modifications to the existing prognostic equations, the scheme could be generalized with more realistic PDFs. With a choice of a different assumed PDF, one would need to map the required moments with the PDF variables and either add or subtract the number of prognostic equations needed for the closure \cite{Larson2002}. The prognostic variables and PDF parameters of the assumed double-$\delta$ PDF are the mean state, the second- and third-order moments of the vertical velocity, and the vertical fluxes of temperature and salt. The more complex the assumed PDF is, the more parameters are needed and higher computational cost required to solve the additional prognostic equations. For brevity, we leave the exploration of different PDFs for future studies and as a first implementation use the double-$\delta$ representation of the fluxes in the two-plume (upwelling/downwelling) framework in the new ocean ADC scheme.

\section{Conclusions}
\label{sec:concl}
In this study we have discussed the assumed distribution higher-order closure (ADC) scheme for the first time to parameterize OSBL turbulence. The advantages of ADC schemes compared to other existing mixing scheme are that it includes energetic constraints and inherently captures both local and non-local mixing across OSBL. By solving a few prognostic equations where the higher-order moments appearing in these equations are diagnosed using mass-flux/PDF concept, ADC provides a closed set of equations for the closure. We have validated ADC parameterization scheme against large eddy simulations for a number of convective test cases. The remarkable agreement between ADC and LES, especially in the mean and turbulent flux profiles is encouraging. We also showed that the ADC scheme is not sensitive to vertical resolutions by testing across resolutions varying order of magnitude and against commonly used OSBL mixing schemes.   

%
%
%
\appendix

\section{Lateral mass exchange: Dissipation and length scales} 
\label{sec_ED}
\begin{figure}[h!]
\centering
\includegraphics[width=7cm]{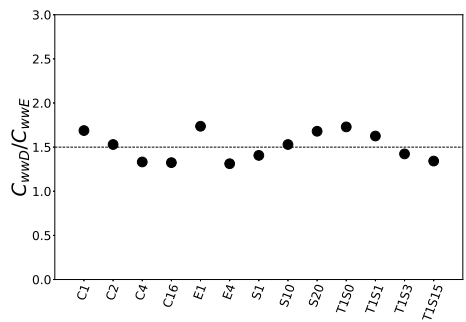}
\caption{Ratio of the dissipation constants diagnosed from LES results for each test case corresponding to maximum convective flux averaged over the final day of the simulations.}
\label{fig:lupld}
\end{figure}

The ADC model uses a more general approach compared to traditional mass-flux parameterizations to represent lateral mass exchange between upwelling and downwelling plumes such that it is also applicable to regimes where both the downwelling and upwelling plumes are turbulent. Lateral mass exchange between upwelling and downwelling plumes is represented through terms of entrainment ($E$) and detrainment ($D$), that appear in the prognostic equations as dissipative terms (Eqs. \ref{eq_dissipation_parameterization}). 
Similar to \citet{Lappen2001c} $E$ and $D$ are parameterized as
\begin{eqnarray}
    E=C_{wwE}\sigma(1-\sigma)M_c/L_u, \\
    D=C_{wwD}\sigma(1-\sigma)M_c/L_d,
\end{eqnarray}
where $C_{wwE}$ and $C_{wwD}$ are dissipation constants due to entrainment and detrainment, respectively, and $L_{d}$ and $L_{u}$ are the turbulent mixing length scales in the downward  and upward direction, respectively. Following the method of \citet{BA86}, \citet{Lappen2001c} provided these length scales such that, for any depth $z$, or level of neutral buoyancy, $L_d$ is the downward distance travelled by a plume due to buoyancy until it overshoots its initial kinetic energy and, similarly, $L_u$ is the upward distance travelled by the plume. $L_d$ and $L_u$ are defined as
\begin{eqnarray}
 L_u= \int_{z}^{z+L_u(z)} [B(z')-B_u(z')]dz' =k(z), \\
 L_d=\int_{z-L_d(z)}^{z} [B_d(z')-B(z')]dz' =k(z),
\end{eqnarray}
where $k=0.5(u'^2+v'^2+w'^2)$ is the turbulent kinetic energy and $B=(g\alpha_T(\theta_0-\overline{\theta})-g\beta_s(S_0-\overline{S}))$ is the average buoyancy surrounding upwelling (downwelling) for estimation of $L_u (L_d)$, not the mean state surrounding \citet{Lappen2001c}. Using a harmonic average of these two length scales, the dissipation length ($L$) that can be used to dissipate all turbulent moments is given as
\begin{linenomath*}
\begin{eqnarray}
   L=\frac{2L_{u}L_{d}}{(L_{u}+L_{d})},
\end{eqnarray}
and the turbulent time scale ($\tau=2k/\varepsilon$ ) is defined as 
\begin{eqnarray}
   \tau=2L/\sqrt{k}.
   \label{eq:tau}
\end{eqnarray}
\end{linenomath*}
The ratio of the dissipation constants are shown to be $C_{wwD}/C_{wwE}=L_{u}/L_{d}=1.5$, corresponding to the depth of maximum convective mass flux (i.e., corresponding to a depth approximately $z/h=0.4$) \cite{Lappen2001c}. We also independently verified this relation using our LES data and found a general agreement with the value of $1.5$ as shown in Figure \ref{fig:lupld}. The dissipation constants used for this study are $C_{wwE}=0.2$ and $C_{wwD}=0.3$.  Since $E$ and $D$ are dynamically parameterized using $\sigma$, $M_c$, and length scales, which are also function of height, the ADC scheme provides physically based parameterizations for the lateral mass exchange between plumes.

\section{Pressure term closure and the horizontal turbulent kinetic energy budgets}
\label{sec:pressure}
Pressure effects appear in the budgets for all the moments that contain velocities (fluxes and variances). Since these turbulent pressure terms are higher-order than the equations they appear in and they depend on the full (non-local) flow field, their effects require a closure \cite{Pearson2019}. These pressure effects can be separated into a pressure-strain, or pressure-scalar term, and the divergence of a pressure-transport term \cite{Launder1975}. The current ADC scheme does not explicitly consider the pressure transport following \citet{Lappen2001a}, but we note there are HOC methods to parameterize the pressure transport or to merge it into advective transport closures \cite{kantha1994}. Closure schemes for the pressure-strain and pressure-scalar terms typically separate pressure effects based on their physical drivers, with the relevant physical mechanisms for convective turbulence being buoyancy effects and turbulence-turbulence interactions.

In this new scheme, the pressure terms in equations  \ref{eq_wtheta_hoc}-\ref{eq_www_hoc} are parameterized following \citet{Canuto2007} as
\begin{linenomath*}
\begin{eqnarray}
    \Pi_{w\theta} &=& -\frac{\overline{w'\theta'}}{\tau_{\theta}} - c_7 \overline{b'\theta'}, \nonumber \\
    \Pi_{ws} &=& -\frac{\overline{w's'}}{\tau_{s}} - c_7 \overline{b's'}, \nonumber \\
    \Pi_{w2} &=& -\frac{2\overline{w'^2} - \overline{u'^2} - \overline{v'^2}}{3\tau_{w2}} - \frac{4}{3}c_{b1}\overline{w'b'}, \nonumber \\
    \Pi_{w3} &=& -\frac{\overline{w'^3}}{\tau_{w3}} - 3c_{b2}\overline{w'^2b'},
    \label{eq:pressure_closure}
\end{eqnarray}
\end{linenomath*}
where the first terms on the right-hand side of equation \ref{eq:pressure_closure} represent the return-to-isotropy process proposed by \citet{Rotta1951}, which removes fluxes and makes the turbulent kinetic energy (TKE) more isotropic, and the second term represents buoyancy-driven pressure effects \cite{Canuto2007}. 
The above closures for pressure terms expand on the closures used in the \citet{Lappen2001a} ADC scheme by including buoyancy terms in the closures for the tracer flux and $\overline{w'^3}$ budgets ($c_7$ and $c_{b2}$ terms respectively).

The return-to-isotropy term in the $\overline{w'^2}$ budget contains the horizontal TKE components, which must be diagnosed from their own budgets. In the convective test cases here, there is no shear production, so these budgets are
\begin{linenomath*}
\begin{eqnarray}
      \frac{\partial \overline{u'^2}}{\partial t} &=& - \frac{\partial \overline{u'^2w'}}{\partial z} + \Pi_{u2} - \frac{2}{3}\varepsilon, \nonumber \\
      \frac{\partial \overline{v'^2}}{\partial t} &=& - \frac{\partial \overline{v'^2w'}}{\partial z} + \Pi_{v2} - \frac{2}{3}\varepsilon,
\label{eq:ADC_uu_budget} 
\end{eqnarray}
\end{linenomath*}
where the dissipation ($\varepsilon$) is estimated as $\varepsilon = k^{3/2}/(c_{\varepsilon}L)$ and $L$ is a length scale described in Appendix \ref{sec_ED}. 

The pressure terms in horizontal TKE budgets are analogous to equation (\ref{eq:pressure_closure}), such that
\begin{linenomath*}
\begin{eqnarray}
    \Pi_{u2} &=& -\frac{2\overline{u'^2} - \overline{v'^2} - \overline{w'^2}}{3\tau_{w2}} + \frac{2}{3}c_{b1}\overline{w'b'}, \nonumber \\
    \Pi_{v2} &=& -\frac{2\overline{v'^2} - \overline{u'^2} - \overline{w'^2}}{3\tau_{w2}} + \frac{2}{3}c_{b1}\overline{w'b'}.
\end{eqnarray}
\end{linenomath*}
By comparing the pressure-strain closures in the vertical and horizontal TKE budgets, it is apparent that any vertical TKE removed by the pressure-strain closure is redistributed to the horizontal TKE. In the convective scenarios presented here, this conversion from $\overline{w'^2}$ is the only source of horizontal TKE.

The third-order transports of horizontal energy are estimated through a down-gradient parameterization as
\begin{linenomath*}
\begin{equation}
    \overline{u'^2w'} = -\kappa \frac{\partial \overline{u'^2}}{\partial z}, \qquad \overline{v'^2w'} = -\kappa \frac{\partial \overline{v'^2}}{\partial z},
\end{equation}
\end{linenomath*}
where $\kappa=C_{mom}L\sqrt{k}$ represents the diffusivity resulting from small-scale mixing. The timescales for the return-to-isotropy terms are 
$\tau_{w3}=\tau/c_{mom-w3}$, $\tau_{\theta}=\tau/c_{slow\theta}$, and $\tau_{w2}=\tau/c_{slow} $. The constants in the above equations are prescribed as $c_\varepsilon=10$, $c_7=0.33$, $c_{b1}=0.5$, $c_{b2}=0.1$, $C_{mom}=0.5$, $c_{mom-w3}=7.1$, $c_{slow\theta}=2.0$, and $c_{slow}=1.38$. Salinity closure parameters are identical to their temperature counterparts.

\section{Sub-plume scale parameterization}
\label{sec:sps}
Small-scale motions in the ADC model, specifically near the surface and for the inversion that is not captured efficiently by the plume-scale fluxes, are captured through sub-plume scale parameterizations. As shown in the prognostic equations in section \ref{sec:methods}, sub-plume scale effects are included in source/sink terms and are assumed to be down gradient. The sub-plume contribution are those proposed by \citet{Lappen2001a,Lappen2001c}
\begin{linenomath*}
\begin{eqnarray}
 SPS_w = \frac{1}{\sigma}\frac{\partial \left[ \sigma \overline{w'^2}_{sps, d}\right]}{\partial z}-\frac{1}{(1-\sigma)}\frac{\partial\left[ (1-\sigma)\overline{w'^2}_{sps,u} \right]}{\partial z},
\end{eqnarray}
\end{linenomath*}
\begin{linenomath*}
\begin{eqnarray}
 SPS_{\theta} = \frac{1}{\sigma}\frac{\partial \left[ \sigma \overline{w'\theta'}_{sps,d}\right]}{\partial z}-\frac{1}{(1-\sigma)}\frac{\partial\left[ (1-\sigma)\overline{w'\theta'}_{sps,u} \right]}{\partial z},
\end{eqnarray}
\end{linenomath*}
where $\overline{w'\theta'}_{sps,d}$ and $\overline{w'\theta'}_{sps,u}$ represent vertical fluxes of temperature due to sub-plume scale (local) mixing in the downwelling and upwelling plume, respectively. These fluxes are estimated through a down-gradient closure 
\begin{linenomath*}
\begin{equation}
  \overline{w'\theta'}_{sps, i}= - K_{h,i} \frac{\partial \overline{\theta}}{\partial z},
  \label{eq_sps_flux}
\end{equation}
\end{linenomath*}
where $K_{h,i}$ is the sub-plume scale eddy diffusivity with $i$ representing $u$ or $d$. This diffusivity is defined as $K_{h,i}=(1+2l_{sps,i}/dz)K_{m,i}$, where $K_{m,i}=0.1l_{sps,i}(e_{sps,i})^{1/2}$ is a sub-plume scale momentum diffusivity. We assume that the sub-plume length scale $l_{sps,i}$ is equivalent to the vertical grid size $dz$ for unstable stratification and $l_{sps,i}=0.76(e_{sps,i}^{3/2}/B)^{1/2}$ for stable stratification. The variable $e_{sps,i}$ is the sub-plume scale turbulent kinetic energy, which is assumed to be isotropic, hence $\overline{w'^2}_{sps, i}=\frac{2}{3}e_{sps,i}$. The scheme solves a prognostic equation for the sub-plume scale energy in each plume
\begin{linenomath*}
\begin{eqnarray}
 \frac{\partial e_{sps,i}}{\partial t}&=& g(\alpha_T \overline{w'\theta'}_{sps,i}-\beta_S \overline{w's'}_{sps,i})\nonumber\\
 &&+\frac{\partial (2K_m\partial e_{sps,i}/\partial z)}{\partial z}-\varepsilon_{sps,i}+\varepsilon, \label{eq:sps_tke_tendency}
\end{eqnarray}
\end{linenomath*}
where $\varepsilon_{sps,i}=C_i e_{sps,i}^{3/2}/l_{sps,i}$ is the sub-plume scale dissipation, $\varepsilon$ is the dissipation of plume-scale vertical kinetic energy through entrainment and detrainment, which acts as a source for sub-plume scale energy, and $C_i=0.19+0.51 l_{sps,i}/dz$. 
 \begin{figure}[h]
\centering
\includegraphics[width=10cm]{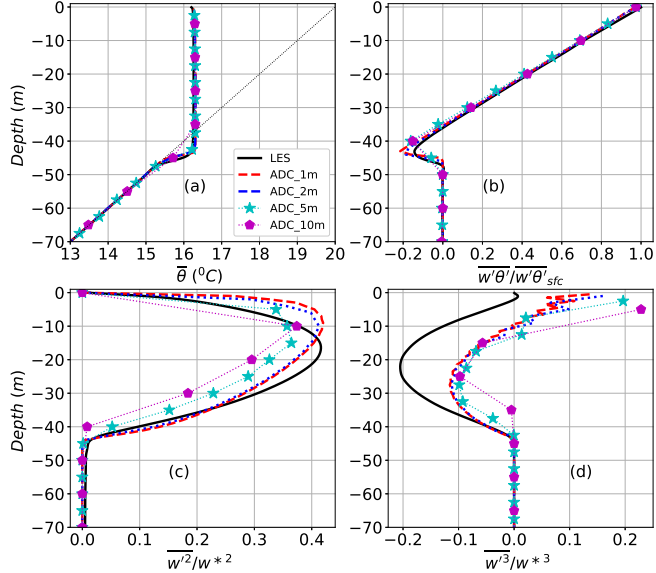}
\caption{Mean profiles of (a) temperature $(\overline{\theta})$, (b) heat flux $(\overline{w'\theta'})$, (c) vertical velocity variance $(\overline{w'^2})$, and (d) the third-order moment of vertical velocity $(\overline{w'^3})$ for LES (black solid lines) and the ADC model with a $1$~m resolution (red dashed lines), $2$~m resolution (blue dashed lines), $5$~m resolution (dotted line with cyan star markers), and $10$~m resolution (dotted line with magenta pentagon markers) for the strongest heat flux case (C16).}
\label{fig:c16_all}
\end{figure}
\section{Sensitivity to Vertical resolution and time-step}
\label{sec:reso_sensitivity}

For an OSBL parameterization to be widely and reliably usable within a GCM, its results must be insensitive to model vertical resolution and time step size. To test the vertical resolution sensitivity of the ADC model, we ran each case at four different vertical resolutions: 1~m, 2~m, 5~m, and 10~m, with the latter being comparable to typical near surface resolution in global ocean GCMs. 

\begin{figure}[h]
\centering
\includegraphics[width=9cm]{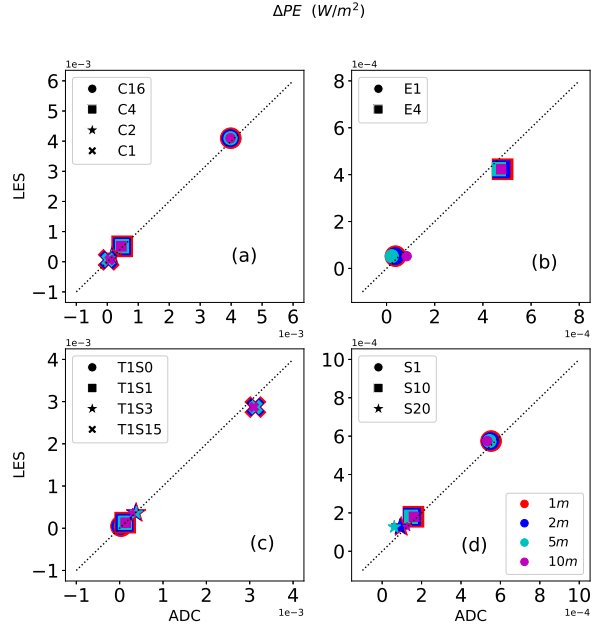}
\caption{Rate of change in depth integrated potential energy for LES (vertical axis) vs the ADC model (horizontal axis) for ADC resolutions of $1$~m (red), $2$~m (blue), $5$~m (cyan), and $10$~m (magenta) for (a) surface heat flux (C\#), (b) surface salinity flux (E\#), (c) initial stratification (S\#), and (d) combined surface heat and salinity flux (T1S\#) cases. Colors indicate resolution and markers indicate different forcing/stratification. The dotted black line shows a one-to-one comparison of LES and the ADC model. Different marker sizes are used for clearer visibility of collocated markers.}
\label{fig:pe_change_adc_les}
\end{figure}

Figure \ref{fig:c16_all} shows mean profiles of temperature $(\overline{\theta})$, heat flux $(\overline{w'\theta'})$, vertical velocity variance $(\overline{w'^2})$, and the third-order moment of vertical velocity $(\overline{w'^3})$ from LES and from the ADC model with different vertical resolutions for a representative test case (C16, similar results are seen for all other test cases). The ADC temperature, heat flux, and vertical velocity variance profiles agree well with LES across all resolutions. The $\overline{w'^3}$ magnitudes are consistently smaller in ADC than in LES regardless of the resolution. We note that accurate estimation of $\overline{w'^3}$, especially at the surface is extremely difficult in any closure model and many higher-order models impose artificial realizability constraints \cite{Andre1978}. However, ADC estimated $\overline{w'^3}$ is qualitatively comparable to LES. Despite an order of magnitude change in the vertical resolution, the ADC model results are mostly insensitive to the resolution used, with the exception of small deviations in the depth of the maximum vertical velocity variance, which is consistent across all resolutions. This suggests the utility using our ADC scheme in GCMs over the commonly used KPP parameterizations scheme, which shows a high sensitivity to vertical resolution \cite{vanRoekel2018, Li2019} (also shown in Figures \ref{fig:pe_ke_kpp_adc}-\ref{fig:re_pe_all}). 
To demonstrate the insensitivity of the ADC scheme to vertical resolutions in terms of energetics, we show the rate of change in the depth integrated potential energy from its initial profile (equation \ref{eq:deltape}) for all the test cases and across all the resolutions in Figure \ref{fig:pe_change_adc_les}. 

\begin{figure}[h!]
\centering
\includegraphics[width=12cm]{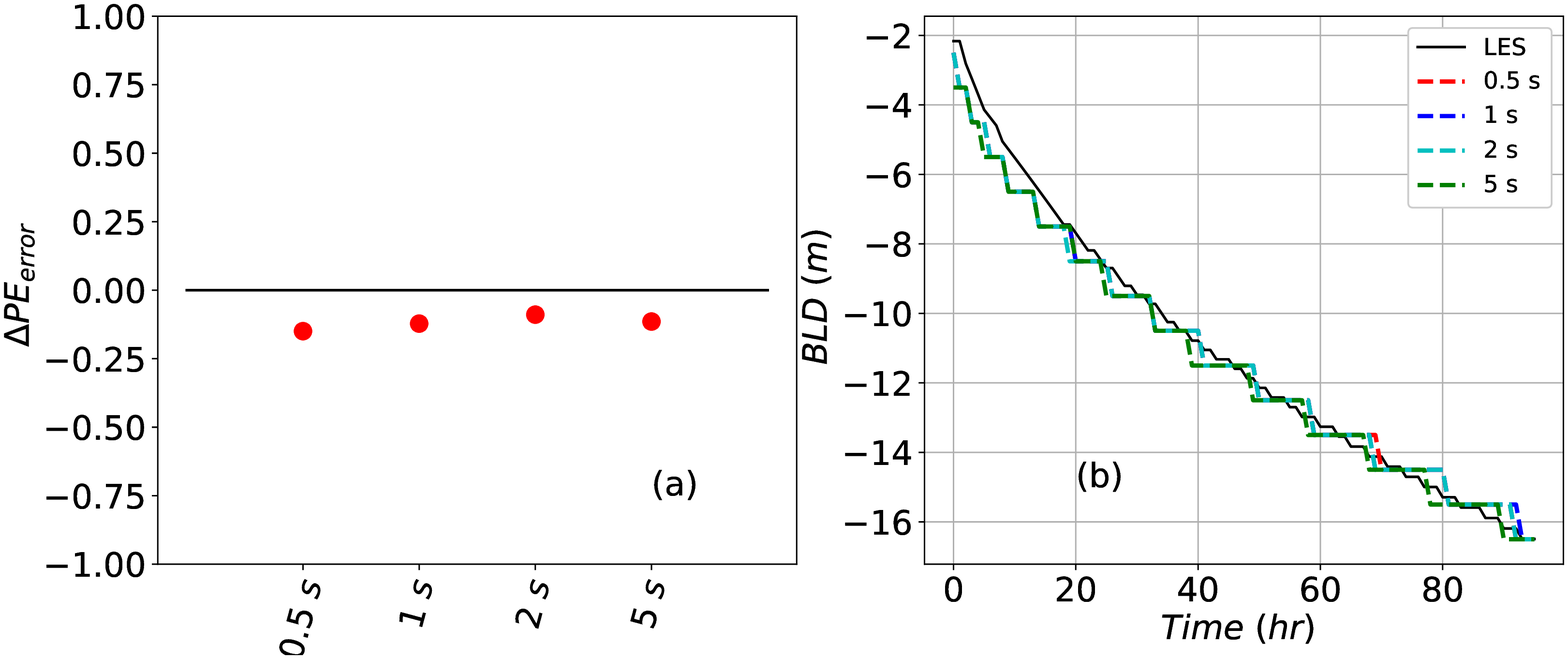}

\caption{(a) Relative difference in the rate of change in depth integrated potential energy between LES and ADC model results while varying the ADC model time step for the C2 test case.  (b) Time evolution of the boundary layer depth for the C2 test case while varying the time step for the LES (black solid lines) and the ADC model with a $0.5$~s time step (red dashed line), a $1$~s time step (blue dashed line), a $2$~s time step (cyan dashed line), and a $5$~s time step (green dashed line).}
\label{fig:dt_adcdt}
\end{figure}

To test the time step sensitivity of the ADC model, we vary the sub-cycle turbulent time step between $0.5$~s, $1$~s, $2$~s, and $5$~s, while keeping the vertical resolution at $1$~m. To quantify sensitivity, we compute the relative difference in the rate of change in integrated potential energy between the LES and ADC model (equation \ref{eq:RE_deltaPE}). Figure \ref{fig:dt_adcdt}(a) shows the relative difference of the change of depth integrated potential energy for the C2 case. Similar results are seen in all cases, since they are all free convective turbulent mixing cases. Only a very weak sensitivity to time step is seen with the ADC model. 
To better visualize the insensitivity of the ADC scheme's time evolution to time step size, Figure \ref{fig:dt_adcdt}(b) shows the time evolution of the boundary layer depth (h) as the turbulent time step is varied for the C2 test case, where, again, h is defined as the depth of the maximum in $N^2$. For convective test cases without a mean current, the mixing scheme is insensitive to the GCM time step used in MPAS-Ocean (not shown).\\\\

\acknowledgments
This research was funded as part of the Energy Exascale Earth System Model (E3SM) and Interdisciplinary Research for Arctic Coastal Environments (InteRFACE) project through the Department of Energy, Office of Science, Biological and Environmental Research Earth and Environment Systems Sciences Division, Regional and Global Model Analysis (RGMA), Earth System Model Development (ESMD), MultiSector Dynamics (MSD), and Data Management (DM) programs and was awarded under contract grant 89233218CNA000001 to Triad National Security, LLC (``Triad"). We thank the two anonymous reviewers for their constructive comments to help improve our manuscript.\\\\
The source code for proposed new mixing scheme is available at https://github.com/\\vanroekel/MPAS-Model/tree/ocean/addADCMixing. The data for the study are available at https://doi.org/10.5281/zenodo.6404409 and the scripts used to produce the figures in the manuscript are available at https://github.com/amrapallig/adcAnalysis.


%
%
\bibliography{ADC_1.bib}




\end{document}